\documentclass[reprint,twocolumn,amsmath,amssymb,aps,superscriptaddress]{revtex4-2}

\usepackage{graphicx}
\usepackage{dcolumn}
\usepackage{bm}
\usepackage{graphicx}
\usepackage{amssymb}
\usepackage{epstopdf, epsfig}
\usepackage{yfonts}
\usepackage[colorlinks=true, linkcolor=blue, citecolor=blue, urlcolor=blue]{hyperref}
\usepackage{fontenc}
\usepackage{upgreek}
\usepackage{mathrsfs}
\usepackage{multirow} 
\usepackage{booktabs}
\usepackage{pifont}
\usepackage{amsmath}
\usepackage{bm}
\usepackage{wasysym}
\usepackage{caption}
\usepackage{subcaption}
\usepackage{euscript}
\usepackage{natbib}
\usepackage{lipsum}
\usepackage{xcolor}
\usepackage{cases}

\usepackage{orcidlink}

\usepackage{subcaption}
\usepackage{ragged2e}
\DeclareCaptionJustification{justified}{\justifying}
\begin{document}

\title{Effect of shoaling length on rogue wave occurrence}

\author{Jie Zhang\,\orcidlink{0000-0003-0794-2335}}
\email{jie.zhang@dlut.edu.cn}
\affiliation{State Key Laboratory of Coastal and Offshore Engineering, Dalian University of Technology, Dalian 116023, PR China}

\author{Saulo Mendes\,\orcidlink{0000-0003-2395-781X}}
\email{saulo.dasilvamendes@unige.ch}
\affiliation{Group of Applied Physics, University of Geneva, Rue de l'\'{E}cole de M\'{e}decine 20, 1205 Geneva, Switzerland}
\affiliation{Institute for Environmental Sciences, University of Geneva, Boulevard Carl-Vogt 66, 1205 Geneva, Switzerland}

\author{Michel Benoit\,\orcidlink{0000-0003-4195-2983}}
\email{michel.benoit@edf.fr}
\affiliation{EDF R\&D, Laboratoire National d’Hydraulique et Environnement (LNHE), Chatou, France}
\affiliation{Saint-Venant Hydraulics Laboratory, LHSV, \'{E}cole des Ponts, EDF R\&D, Chatou, France}

\author{J{\'e}r{\^o}me Kasparian\,\orcidlink{0000-0003-2398-3882}}
\email{jerome.kasparian@unige.ch}
\affiliation{Group of Applied Physics, University of Geneva, Rue de l'\'{E}cole de M\'{e}decine 20, 1205 Geneva, Switzerland}
\affiliation{Institute for Environmental Sciences, University of Geneva, Boulevard Carl-Vogt 66, 1205 Geneva, Switzerland}

\begin{abstract}
The impact of shoaling on linear water waves is well-known, but it has only been recently found to significantly amplify both the intensity and frequency of rogue waves in nonlinear irregular wave trains atop coastal shoals. At least qualitatively, this effect has been partially attributed to the ``rapid'' nature of the shoaling process, i.e., shoaling occurs over a distance far shorter than that required for waves to modulate themselves and adapt to the reduced water depth. Through the development of a theoretical model and highly accurate nonlinear simulations, we disentangle the respective effects of the slope length and the slope gradient of a shoal and focus on the slope length to investigate the rapidness of the shoaling process on the evolution of key statistical and spectral sea-state parameters. Provided the shoal slope is 1/10 or steeper, our results indicate that the non-equilibrium dynamics is involved even for rather short shoaling lengths and becomes dominant in the regime of large lengths. When the non-equilibrium dynamics governs the wave evolution, further extending the slope length no longer influences the statistical and spectral measures. Thus, the shoaling effect on rogue waves is mainly driven by the slope magnitude rather than the slope length. Moreover, the simulations show that a higher cut-off frequency of the wave spectrum has a smaller impact on wave statistics than expected for a flat bottom in deep water and that insufficient attenuation of low-frequency waves at the downstream domain boundary has notable influence on wave statistics atop the shoal.
\end{abstract}


\maketitle

\section{INTRODUCTION}\label{sec:intro}

Quantifying the impact of rogue waves on the stability of offshore and coastal structures has become a recent direction of research. An increased frequency of rogue waves in a time series may lead to an excess in wave loads \citep{Adcock2022, Chabchoub2023b, Adcock2023, Swan2023, Li2023}. 
The statistics of water waves define the design envelope for ocean vessels, and the possibility of avoiding or mitigating these extreme waves is important for marine and coastal safety.
The risk of rogue waves in deep water is nowadays rather well-known \citep{BPelinovsky2008, Chabchoub2011, Chabchoub2012, Fedele2016}, but rogue waves are also observed in intermediate \citep{Trulsen2012, Trulsen2012b, Gramstad2013, Viotti2014} and shallow waters~\citep{Sergeeva2011, Pelinovsky2016,Mendes2021, Karmpadakis2022, Didenkulova2023} subject to bathymetry gradients. 
Among the mechanisms associated with rogue wave formation and amplification \citep{Dysthe2008,Adcock2014,Onorato2013}, shoaling leads to the highest recorded excess in kurtosis~\citep{Chabchoub2023}. 

When propagating nearshore, the hydrodynamics of wind-generated surface waves is affected by the relative water depth until wave-breaking becomes dominant \citep{Green1838,Burnside1915,Holthuijsen2007}, and modifies the wave dispersion, wavelength, group velocity, wave height, among others. Several models describe the deformation and transformation of waves encountering bathymetric changes~\citep{Eagleson1956,Shuto1974,Walker1983,Goda1997}, but rogue wave amplification by the non-equilibrium process of shoaling cannot be calculated directly from the shoaling coefficient and the behaviour of regular waves only. 
Indeed, rogue waves seem to respond to dimensionless rather than dimensional parameters such as significant wave height or wavelength. Theoretical, experimental, and numerical studies addressed the amplification of rogue waves in the past years~ \citep{Kashima2014,Ma2015,Ducrozet2017,Moore2019,Zhang2019,Moore2020,Trulsen2020,Adcock2020,Doeleman2021,Ma2021,Chabchoub2021b,Chabchoub2021,Adcock2021a,Adcock2021b,Mori2021,Xu2021,Zhang2021,Trulsen2022,Mendes2022,Mori2023,Benoit2023}. A key component of this amplification is the abruptness of the environmental transition, which often drives physical systems out of equilibrium \citep{Lockwood2001,Sobolev2013,Steinbach2016,Passiatore2022}, thereby generating transient local phenomena until the new equilibrium state is established \citep{Zhang2023}. \citet{Trulsen2018} anticipates that variations in environmental or meteorological conditions over a rapid spatio-temporal scale will in general lead to out-of-equilibrium states and thus to anomalous wave statistics. This type of anomalous statistics is also observed in the generation of irregular fields in wave flumes or basins when the sea-states are forced by random input spectra. In such cases, the kurtosis behaves anomalously within a spatio-temporal scale shorter than the characteristic wave scale \citep{Tang2020}.

Shoaling is characterized by three parameters: the slope gradient, its spatial extension, and the change in depth. These parameters however provide only two degrees of freedom. Here we consider the slope magnitude and the slope length (equivalently called ``shoaling length'' hereafter), from which the depth change can be determined correspondingly. 
To the limit of our knowledge, explicit expressions involving rapid transition or slope magnitude are missing in second-order theories, except in the recent work of~\citet{Mendes2022b} for the latter. Nonetheless, it is often argued that disturbances in wave statistics occur due to the ``abrupt'' nature of the depth transition, which drives the system out of equilibrium. Especially for a submerged step, when wave packets evolving slowly in time and space meet an abrupt depth transition, the modulation between the envelopes of bound and free superharmonics would occur \citep{Adcock2021a}. However, the term ``abruptness'' is generally used in an intuitive manner, without clearly defining whether it refers to a strong bathymetry change, a spatial extension shorter than the wavelength, or a steep slope~\citep{Viotti2014,Adcock2020,Adcock2021c,Lawrence2021a,Draycott2022}. In an engineering context, the term ``abrupt depth transition'' is also used for a variety of natural and man-made bathymetric features \citep{Draycott2022} although their slopes are well below the threshold of 1/3 defined for abruptness by the mild slope equation \citep{Booij1983}. In this context, abruptness rather appears to refer to depth differences. These ambiguities result in confusing and sometimes seemingly contradictory results.

There is widespread agreement that the steepness growth up to the second order is the main driver of out-of-equilibrium phenomena in intermediate water depths. However, experimentation allowing to distinguish between the respective effects of slope magnitude and shoaling length under the same conditions as in the experiments of \citet{Trulsen2020} are still lacking. Here, we address the call of \citet{Trulsen2018} for the quantification and verification of the effect of the shoaling length. We define a shoaling length parameter $\ell$ as the ratio of the length of the shoal to the wavelength. We disentangle the effect on rogue wave probability of this measure of the wave shoaling ``abruptness'' from the effect of slope magnitude. Relying on non-homogeneous second-order theory and numerical simulations using a fully nonlinear potential flow (FNPF) model, we vary the length of the shoal while keeping a fixed slope magnitude and the same sea-state conditions atop the shoal. We show that, unlike expectations, the effect of a shoal on the rogue wave probability amplification is not governed by the shoaling length, thus purely by the slope magnitude, as soon as this length exceeds about half of the peak wavelength.


\section{Shoaling length and non-homogeneous wave theory}\label{sec:theory} 
The present work investigates the effect of the shoaling length by disentangling it from the slope magnitude $|\nabla{h}|$, with $h$ denoting the still water depth ($h > 0$) and $\nabla$ the gradient operator in the horizontal plane. Noting that $\partial h/ \partial x < 0$ in the current setup with a reduction of water depth due to the shoal with increasing $x$, the bottom slope will be denoted as $|\nabla{h}|$ hereafter. We define the shoaling length parameter as $\ell \equiv L/\lambda$, with $L$ denoting the length of the shoal and $\lambda$ the characteristic wavelength. 
In this section, we use the theoretical reasoning from \citet{Mendes2022} to reveal the effect of $\ell$ on anomalous wave statistics after a steep shoal, as it delineates distance ranges from the start of the shoal to the location of the peak kurtosis and would constrain laboratory experiments of shoaling rogue waves.

\subsection{Effect of $\ell$ on exceedance probability of wave heights}

\begin{figure*}
\centering
    \includegraphics[scale=0.8]{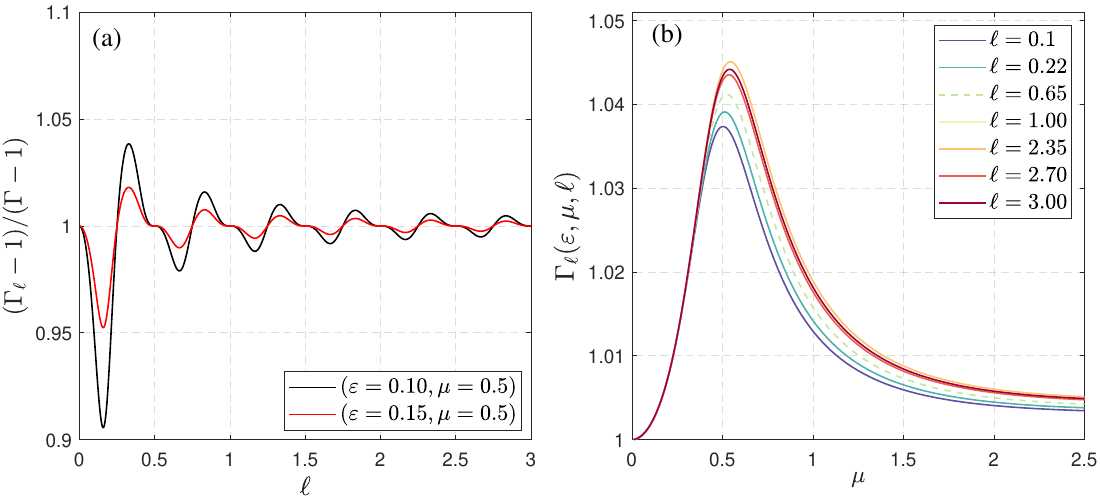}
    \caption{\centering{Effect of the non-dimensional shoaling length parameter $\ell$ on (a) the ratio of the relative non-homogeneous parameter $(\Gamma_{\ell} - 1)/(\Gamma - 1)$ and (b) the evolution of $\Gamma_{\ell}$ as a function of relative water depth $\mu =kh$, with steepness $\varepsilon=0.05$ and $\ell \in [0.1,3.0]$. $\Gamma_{\ell}$} is computed with the formulation of eq.~\eqref{eq:gammaL} in both panels.}
\label{Fig:ell}
\end{figure*}
A non-homogeneous stochastic theory describes the energy redistribution among modes of irregular water waves travelling over a shoal of arbitrary slope magnitude \citep{Mendes2023}. Taking into account these disturbances, we may approximate the exceedance probability $\mathcal{R}$ (for wave height $H$ exceeding $\alpha$ times the significant wave height $H_s$) in a narrow-banded irregular wave train moving past a steep slope through the formula \citep{Mendes2022}:
\begin{equation}
\mathcal{R}(H>\alpha H_{s}) = e^{-2\alpha^{2}/\Gamma} \,\,\, , 
\label{eq:Rayexc0}
\end{equation}
where the non-homogeneous parameter $\Gamma$ is defined as,
\begin{equation}
\Gamma \equiv \frac{ \mathbb{E}[\eta^{2}] }{ \mathscr{E} }  =  \frac{ 1+   \frac{\pi^{2} \varepsilon^{2} }{16}     \, \Tilde{\chi}_{1} }{  1+   \frac{\pi^{2} \varepsilon^{2} }{32}   \, \left( \Tilde{\chi}_{1}  + \chi_{1} \right) +\check{\mathscr{E}}_{p2}   } \,\, , 
\label{eq:Rayexc}
\end{equation}
with: 
\begin{eqnarray}
\Tilde{\chi}_{1} &=& \frac{(3 - \sigma^2)^2} {\sigma^6}  \,\, , 
\label{eq:Rayexc_para}
\\
\nonumber
\chi_{1} &=& \frac{9\, \textrm{cosh}(2\mu) }{\textrm{sinh}^{6} \mu} =\frac{9}{\sigma^6}(1-\sigma^2)(1-\sigma^4)\,\, , \,\, \sigma \equiv \tanh\mu \, .
\end{eqnarray}
$\mathbb{E}[\eta^{2}]$ denotes the ensemble average of $\eta^2$, and is approximated by the variance $\langle \eta^2 \rangle$ of the surface elevation $\eta$ in a discrete time series. $\mathscr{E}$ is defined as the total mechanical energy averaged over one wavelength \citep[see][for instance]{Dalrymple984} and divided by $\rho g$ where $\rho$ is the water density and $g$ the acceleration due to gravity. For linear regular waves, we have $\mathbb{E}[\eta^{2}]= \mathscr{E}=a^{2}/2$ \citep{Airy1845}, and thus $\Gamma =1$. The irregular wave steepness $\varepsilon \equiv (\sqrt{2}/\pi) k_{p}H_{s}$ is computed from the ratio between the significant wave height and the peak wavelength $\lambda_p=2\pi/k_p$ with $k_p$ the peak wavenumber, $\mu \equiv k_{p}h$ is the relative water depth, and $\check{\mathscr{E}}_{p2} $ is the non-dimensional net change in potential energy due to the set-down. The potential energy is a function of the slope magnitude $|\nabla h|$ over a shoaling zone, and of the net change of the potential energy due to set-down, $\check{\mathscr{E}}_{p2}$. Consistent with deep water pre-shoal conditions, the latter is approximated as \citep{Mendes2022b}:
\begin{eqnarray}
\nonumber
\check{\mathscr{E}}_{p2} &\approx & \frac{5\varepsilon^{2}}{\mu^2} \left[   \frac{ 1 }{125 |\nabla h|} -  |\nabla h| \left( 1 - |\nabla h|    \right)    \right]  
\\
&\equiv & \frac{\varepsilon^{2}}{\mu^2}  \, f(|\nabla h|) \,\,\, , \,\,\, 1/100 \leqslant |\nabla h| \leqslant 1/2 \quad .
\label{eq:ep2}
\end{eqnarray} 
At steep slopes (in the vicinity of $|\nabla h| \sim 1/2$) the effect of $\check{\mathscr{E}}_{p2}$ saturates and the term $\check{\mathscr{E}}_{p2} $ can be safely neglected in the model of eq.~(\ref{eq:Rayexc}).

To examine the effect of the shoaling length on wave statistics, we consider a 2D Cartesian coordinate system $(x,z)$ with its $z$-axis origin located at the still water level (SWL) and pointing upward. Then, we calculate the energy $\mathscr{E}$ via an average over an arbitrary length by changing the interval of integration in eq.~(2.4) of \citet{Mendes2022} from one wavelength to the length of the shoal $L$: 
\begin{eqnarray}
\nonumber
 \mathscr{E} \, &\equiv & \mathscr{E}_{p} + \mathscr{E}_{k} =   \frac{1}{2L} \int_{0}^{L}  \eta^{2} (x,t) \, \textrm{d}x  
\\ 
 &+& \frac{1}{2gL } \int_{0}^{L} \int_{-h(x)}^{0} \left[ \left( \frac{\partial \Phi}{\partial x} \right)^{2}  + \left( \frac{\partial \Phi}{\partial z} \right)^{2} \right] \, \textrm{d}z \, \textrm{d}x \,\, ,
\label{eq:energyX0}
\end{eqnarray}
where $\Phi$ is the velocity potential. Note that for $\ell <1$, the integration is performed over less than a wavelength, leading to spurious oscillations. Without loss of generality, in order to simplify the algebra otherwise fully described in section 3 and appendix B of \citet{Mendes2022}, we use a monochromatic-like wave formalism. The latter is an asymptotically valid approximation to the algebraic description of $\Gamma$ in the limit of a large number of wave components because the terms containing the wave amplitude (the significant wave height) are factored out in eq.~(\ref{eq:Rayexc}), see for instance appendices D3-D4 of \citet{BMendes2020} and section 5.1 of \citet{Boccotti2000} for asymptotic computations.
For a second-order Stokes solution, we have the following velocity potential \citep{Dingemans1997,Mendes2022}:
\begin{eqnarray}
 \Phi = \frac{a\omega }{k} \left[  \frac{\cosh{\varphi} }{\sinh{ \mu }} \sin{\phi} + \left(  \frac{3\pi \varepsilon}{8}  \right) \frac{\cosh{(2\varphi)} }{\sinh^{4}{\mu}} \sin{(2\phi)}  \right] \quad , 
\label{eq:seteq0}
\end{eqnarray}
with notations $\varphi \equiv k (z+h)$ and $\phi \equiv kx - \omega t$ and $a$ being the wave amplitude. This leads to the FSE solution \citep{Dingemans1997}:
\begin{equation}
 \eta  =   a \left[ \cos{\phi} +    \left( \frac{ \pi \varepsilon  }{ 4  } \right) \sqrt{\tilde{\chi}_{1}} \cos{(2\phi)} \right]  \quad .
\label{eq:seteq}
\end{equation}
Furthermore, the horizontal and vertical velocity components have the form:
\begin{eqnarray}
  \frac{\partial \Phi}{\partial x} &=&  a\omega \left[ \frac{ \cosh{\varphi} }{ \sinh{ \mu } } \cos{\phi}  + \left(  \frac{3\pi \varepsilon}{4}  \right) \frac{\cosh{(2\varphi)} }{\sinh^{4}{ \mu }} \cos{(2\phi)} \right] \quad  ,
   \label{eq:seteqx}
 \\
  \frac{\partial \Phi}{\partial z} &=& a\omega \left[ \frac{ \sinh{\varphi} }{ \sinh{ \mu } } \, \sin{\phi} \,  + \left(  \frac{3\pi \varepsilon}{4}  \right) \frac{ \sinh{(2\varphi)} }{ \sinh^{4}{ \mu } } \, \sin{(2\phi)} \, \right] \quad .
\end{eqnarray}
Hence, as detailed in appendix~\ref{sec:append_abruptE}, it can be shown that the shoaling length corrections to the variance and energy are small oscillations over $\ell$ (see figure~\ref{Fig:ell}(a)). We obtain the expression of the non-homogeneous parameter $\Gamma_{\ell}$:
\begin{equation}
\Gamma_{\ell} \approx \frac{ 1+   \frac{\pi^{2} \varepsilon^{2} }{16}     \, \Tilde{\chi}_{1} \mathcal{F}(\ell) }{  1+   \frac{\pi^{2} \varepsilon^{2} }{32}   \, \left( \Tilde{\chi}_{1}  + \chi_{1} \right) \mathcal{F}(\ell)   } \quad ,
\label{eq:gammaL}
\end{equation}
where we have defined a correction due to the shoaling length  to the leading order as:
\begin{equation}
 \mathcal{F}(\ell) = \frac{ 1 + \frac{ \sin{(8\pi \ell)} }{8\pi \ell} }{ 1 + \frac{ \sin{(4\pi \ell)} }{4\pi \ell}  } \quad , \quad \forall \,\,\, \ell \geqslant 0 \quad .
 \label{eq:gammaL2}
\end{equation}
Note that for a broad-banded sea, we would have the above equation rewritten with the scaling $\varepsilon \rightarrow  \mathfrak{S} \varepsilon$, where $\mathfrak{S}$ is the mean vertical asymmetry between wave crests and troughs. It is easily verified that $|\mathcal{F}(\ell) - 1|< 0.45$ for all $\ell$, and for $\ell > 0.5$ we have an even lower local bound $|\mathcal{F}(\ell) - 1|< 0.2$. To illustrate the minor effect of $\ell$ on the parameter $\Gamma$, figure~\ref{Fig:ell}(a) shows the ratio $(\Gamma_{\ell}-1)/(\Gamma-1)$ as a measure of the relative differences between eqs.~\eqref{eq:Rayexc} and \eqref{eq:gammaL}. Within the range from linear to second-order Stokes theory for maximum steepness of $\varepsilon \leqslant 1/7$, it shows that $\ell$ induces very minor effects on $\Gamma_{\ell}$ for $\ell >1$, and consequently on the wave statistics. For $\ell<1$, these oscillations are larger but are of little consequence for the non-homogeneous parameter $\Gamma_{\ell}-1$, since a finite spectral width and random phases in eq.~\eqref{eq:energyX0} would further reduce the amplitude of the oscillations. For instance, summing up over a phase range $[0,\pi/2]$ reduces the oscillation amplitude by 30\%. As a result, oscillations in $\Gamma_{\ell}-1$ stay below $10$\% for all $\ell$ and of 0.5\% for $\Gamma-1$, such that the effect of shoaling length $\ell$ on abnormal statistics is very small. Figure~\ref{Fig:ell}(b) shows the evolution of $\Gamma_{\ell}$ as a function of $\mu$, with $\varepsilon=0.12$ and $\ell\in[0.1, 3]$. It is noticed that the differences among $\Gamma_{\ell}$ with various slope lengths are smaller than those due to change of relative water depth. Interestingly, unlike expectations from \citet{Trulsen2018}, our theoretical predictions show that a shorter shoaling length (i.e., $\ell < 1$) leads to a decrease (although small) in wave statistics at fixed steep slopes. 

\subsection{Effect of $\ell$ on excess kurtosis} 

\begin{figure*}
\centering
    \includegraphics[width=0.85\textwidth]{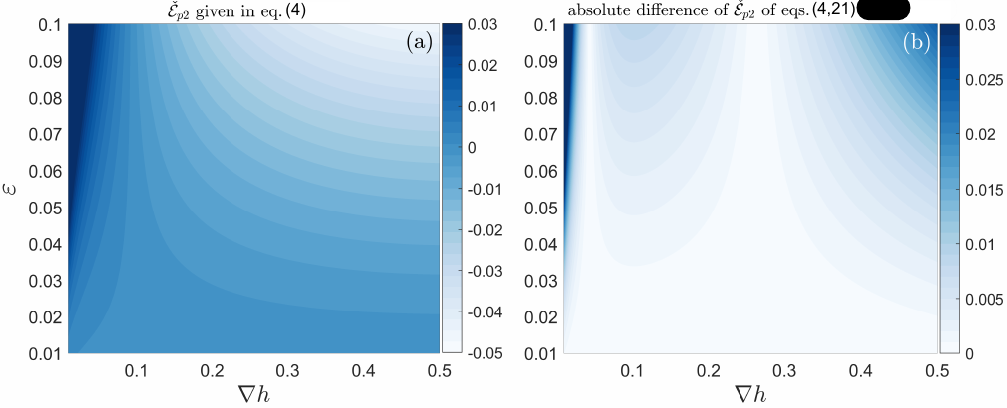}
    \caption{Panel (a): the net change of the potential energy due to set-down $\check{\mathscr{E}}_{p2}$ given in eq.~(\ref{eq:ep2}) for $\mu = 1/2$; and (b): the absolute difference of $\check{\mathscr{E}}_{p2}$ between the formulation of eq.~(\ref{eq:ep2}) and its approximation eq.~(\ref{eq:ep3}) which factors out the slope effect.}
\label{fig:approxslope}
\end{figure*}
The occurrence probability of rogue waves is often investigated through the third and fourth standardized moments (skewness and kurtosis, denoted as $\lambda_3$ and $\lambda_4$, respectively) of FSE as proxies. The n-th standardized moment of a random variable $X$ with zero mean is defined as: 
\begin{equation}
\label{eq:skew_kurt}
    \lambda_n(X) = \frac{\mathbb{E}\left[ X^n \right]}{\left(\mathbb{E}\left[ X^2 \right]\right)^{n/2}},
\end{equation}
where $\mathbb{E}[X]$ denotes the expectancy of $X$. Skewness serves as an indicator of wave asymmetry in the vertical direction resulting from high-order bound harmonics, and kurtosis is positively related to the occurrence frequency of extreme values. We denote the excess kurtosis $\hat{\lambda}_4 \equiv \lambda_4 - 3$. For $X$ following a Gaussian distribution, $\lambda_3(X)=0$ and $\lambda_4(X)=3$ (i.e. $\hat{\lambda}_4 =0$).

To obtain a quantitative prediction for the excess kurtosis of FSE based on the formulae derived in the previous section, we take advantage of an effective theory connecting the non-homogeneous parameter $\Gamma$ and the kurtosis relative to its pre-shoal value in deeper relative waters $\mu_{0}$ (the subscript 0 is used henceforward to denote pre-shoal conditions), namely eq.~(3.3) of \citet{Mendes2023}:
\begin{eqnarray}
\nonumber
 \hat{\lambda}_{4} (\Gamma)&\equiv&\lambda_{4} (\Gamma)-3 
 \\ \nonumber
 &\approx& \frac{1}{9} \Bigg[ e^{ 8 \left( 1 - \frac{ 1 }{ \mathfrak{S}^{2}\Gamma } \right) }- e^{ 8 \Big( 1 - \frac{ 1 }{ \mathfrak{S}^{2}_{_{0}}\Gamma_{_{0}} } \Big) }\Bigg]  
 \\
 &\xrightarrow{ \mu_{0} \rightarrow \infty } & \frac{1}{9} \left[ e^{ 8 \left( 1 - \frac{ 1 }{ \mathfrak{S}^{2}\Gamma } \right) }- 1 \right].
\label{eq:Mori3X}
\end{eqnarray}
The above formulation anticipates that, although $\Gamma_{\ell}-1$ deviates at most $10\%$ from $\Gamma-1$, the exponential character of the excess kurtosis in terms of $\Gamma$ implies a threefold deviation in the excess kurtosis due to the shoaling length for $\ell \ll 1$.
Having the exact values of the variables $(\varepsilon,\mu)$ prior and atop the shoal and $\ell$, we can compute the change in excess kurtosis between the two flat bottom regions. Moreover, the vertical asymmetry for the typical conditions of the simulated wave evolution (see section \ref{sec:analysis}) will be \citep{Mendes2023}:
\begin{equation}
\mathfrak{S}(\alpha = 2) \approx \frac{ (2+6\varepsilon_{\ast})(7+3\varepsilon_{\ast})  }{ 6 (2+3\varepsilon_{\ast})  } \, ,
\label{eq:Gnew}
\end{equation}
where,
\begin{equation}
    \varepsilon_{\ast} \approx \frac{\pi \varepsilon}{3\sqrt{2}}\,  \Big[1- \nu \sqrt{2} + 6 \, \nu^{2} \Big] \Big(\tilde{\chi}_{0}  + \frac{\sqrt{\tilde{\chi}_{1}}}{2} \Big). 
\end{equation}
It can be further approximated for a typical JONSWAP spectral measure of bandwidth $\nu \sim 0.4$ \citep{Higgins1975} and $\mu \sim 1$, leading to $\varepsilon_{\ast} \sim (3/4) \varepsilon$ and thus $\mathfrak{S} \gtrsim (7/6) (1+\varepsilon)$. In order to provide a manageable explicit expression for the excess kurtosis, the non-homogeneous correction is reformulated through a simple Taylor expansion of eq.~\eqref{eq:Rayexc}:
\begin{eqnarray}
\Gamma \approx 1 + \left( \frac{\pi \varepsilon}{4} \right)^{2}  \left( \frac{ \tilde{\chi}_{1} - \chi_{1} }{2}\right) - \check{\mathscr{E}}_{p2} \,\, \, .
\label{eq:GammaApprox}
\end{eqnarray}
Now, the closed forms for the trigonometric coefficients in eq.~(\ref{eq:Rayexc_para}) lead to $\Tilde{\chi}_{1}/\chi_{1} \xrightarrow{\mu \leqslant 1} 1 + (3/2) \mu^{4}$ as the water depth decreases towards shallow waters, so that we can rewrite the correction $\Gamma$ in eq.~(\ref{eq:GammaApprox}) as follows:
\begin{eqnarray}
\nonumber
 \Gamma & \approx  & 1 + \left( \frac{\pi \varepsilon}{4} \right)^{2}   \frac{ \chi_{1}}{2} \cdot \frac
{3}{2} \mu^{4}- \frac{\varepsilon^{2}}{4\mu^{2}} \cdot 4f(|\nabla h|) \,\, ,
\\
&\approx &  1 + \frac{\varepsilon^{2}}{4\mu^{2}} \left[ 2\pi^{2}  -  4f(|\nabla h|)  \right]  \,\, .
\label{eq:GammaApprox2}
\end{eqnarray}
As described in section 3.4 of \citet{Mendes2022}, the effect of asymmetry can be parameterized for second-order waves at the region of the peak in amplification $\mu \in [0.5, 1.0]$:
\begin{equation}
\mathfrak{S}^{2}\Gamma \sim \Gamma^{6} \quad .
\label{eq:Gamma6}
\end{equation}
Hence, except for very small values of steepness ($\varepsilon < 1/50$), the argument of the exponential function in the last part of eq.~(\ref{eq:Mori3X}) can be expanded by using eqs.~(\ref{eq:Gamma6}) and (\ref{eq:GammaApprox}):
\begin{eqnarray}
\hspace{-0.3cm}
\nonumber
 1 - \frac{ 1 }{ \mathfrak{S}^{2}\Gamma }   & \approx &   1 - \left\{ 1 + \frac{\varepsilon^{2}}{4\mu^{2}}  \big[ 2 \pi^{2} -  4f(|\nabla h|) \big]   \right\}^{-6}  \,\,  ,
 \\
 & \approx &    \frac{3\varepsilon^{2}}{2\mu^{2}}   \big[ 2 \pi^{2} -  4f(|\nabla h|) \big] \equiv \frac{3\pi^2 \varepsilon^{2}}{\mu^{2}} -6 \check{\mathscr{E}}_{p2}  .
\end{eqnarray}
Considering that the slope-independent background excess kurtosis after saturation of the slope effect, $\hat{\lambda}_{4 \, , \, b}$, amounts to $ \hat{\lambda}_{4 \, , \, b} =  (e^{ 24 \pi^{2} \varepsilon^{2}/\mu^2}- 1)/9 $, algebraic manipulation leads to:
\begin{eqnarray}
\nonumber
\hat{\lambda}_{4} &=& \frac{ e^{ -48 \check{\mathscr{E}}_{p2} } }{9} \left[ e^{ 24 \pi^{2} \varepsilon^{2}/\mu^2  }- 1 \right] + \frac{1}{9} \left( e^{ -48 \check{\mathscr{E}}_{p2}} -1 \right) 
\\
&=& \hat{\lambda}_{4 \, , \, b} \,   e^{ -48 \check{\mathscr{E}}_{p2} } + \mathcal{O}(10\check{\mathscr{E}}_{p2}) \,\, .
\end{eqnarray}
The second term is one order of magnitude smaller than the background kurtosis $\hat{\lambda}_{4 \, , \, b}$ since $\check{\mathscr{E}}_{p2} \sim 10^{-2}$, thus we can overlook it for the sake of reaching a simplified formulation. Furthermore, the net potential energy form given by eq.~(\ref{eq:ep2}) can be numerically approximated as follows:
\begin{equation}
\check{\mathscr{E}}_{p2} \approx \frac{\varepsilon^{2}}{4\mu^2} \left( 7 - 20 \sqrt{ |\nabla h| }   \right)    \quad .
\label{eq:ep3}
\end{equation}
Figure~\ref{fig:approxslope} displays the comparison of eq.~(\ref{eq:ep2}) and eq.~(\ref{eq:ep3}) in the range $|\nabla h| \in [1/100,1/2]$ and $\varepsilon \in [1/100, 1/10]$ covering the wave conditions in the present study, showing that the approximated expression eq.~(\ref{eq:ep3}) represents well the original net potential energy formulation over a wide range of slope magnitudes. Using $e^{ -48 \check{\mathscr{E}}_{p2}} \approx  1 - 48 \check{\mathscr{E}}_{p2}$ and noticing that for small amplitude waves with steepness of the order of $\varepsilon = 1/20$ second-order waves fulfil $400 \varepsilon^{2} \lesssim 1$, the background kurtosis can be approximated as $10\pi^{2}\varepsilon^{2}/\mu^2$ and thus the excess kurtosis up to second-order in steepness reads:
\begin{equation}
\hat{\lambda}_{4} \approx 20 \pi^{2}   \left( \frac{\varepsilon}{\mu} \right)^{2}  \sqrt{|\nabla h|} 
 \quad ,\, \textrm{for }\, \mu_{0} \gg 1  \,\,\, \textrm{and} \,\,\, \varepsilon \lesssim 1/20 \quad  . 
\label{eq:Mori3Xapprox}
\end{equation}
For the general case of non-Gaussian pre-shoal condition, the above equation should adopt $[(\varepsilon/\mu)^2 - (\varepsilon_{0}/\mu_{0})^2]$ instead of $(\varepsilon/\mu)^2$.
Noteworthy, due to repeated Taylor expansions, eq.~(\ref{eq:Mori3Xapprox}) will not be able to capture all features of the spatial evolution of the excess kurtosis, which are better described by eq.~(\ref{eq:Mori3X}). As such, eq.~(\ref{eq:Mori3Xapprox}) is recommended to compute the peak in maximum excess kurtosis atop the shoal relative to the pre-shoal condition. Furthermore, this approximation and the computation of $\Gamma$ in eq.~(\ref{eq:Rayexc}) is restricted to both linear and second-order waves, as we have assumed that $H_{s}(x) \ll h(x)$. In the next sections, our theory and its inference will be validated by confronting it with fully nonlinear simulation results.

\section{Numerical method: fully nonlinear potential flow (FNPF) model\label{sec:w3d}} 

\begin{figure*}
\centering
    \includegraphics[width=0.65\textwidth]{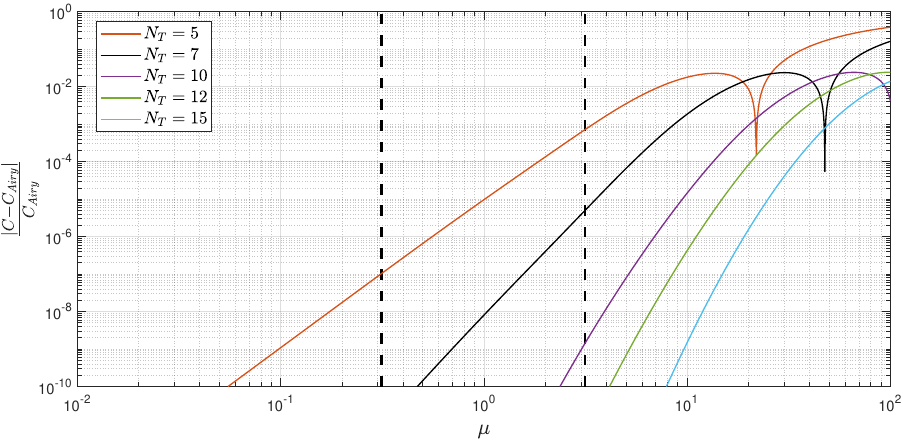}
    \caption{\centering{Relative error of the phase velocity predicted by the linear wave version of the W3D model with different values of $N_T$ in comparison with the analytical solution of Airy linear wave theory. Vertical dashed lines denote the generally adopted shallow- and deep-water limits, $\mu= kh=\pi/10$ and $\mu=\pi$, respectively.}}
\label{Fig:Disper}
\end{figure*}
In order to investigate the effects related to the shoaling length parameter of bottom change and to validate the predictions of the non-homogeneous second-order theory, we performed fully nonlinear numerical simulations within the framework of two-dimensional (2D) FNPF theory. The FNPF theory assumes that the fluid is inviscid and incompressible, and the flow is irrotational. In addition, in this study, the free surface tension is ignored, and the seabed elevation is fixed in time. The water wave problem can then be formulated in terms of the velocity potential $\Phi(x,z,t)$ and the FSE $\eta(x,t)$: 
\begin{eqnarray}
    \hspace{-0.5cm}
    \label{eq:system1.1} \frac{\partial^2 \Phi}{\partial x^2}+\frac{\partial^2 \Phi}{\partial z^2} &=& 0 \text{, in } -h  \le z \le \eta,
    \\
    \hspace{-0.5cm}
    \label{eq:system1.2} \frac{\partial \eta}{\partial t}+\frac{\partial \Phi}{\partial x}\frac{\partial \eta}{\partial x}  &-&\frac{\partial \Phi}{\partial z} =0 \text{, on } z = \eta,
    \\
    \hspace{-0.5cm}
    \label{eq:system1.3}  \frac{1}{2}\left[\left(\frac{\partial \Phi}{\partial x}\right)^2 + \left(\frac{\partial \Phi}{\partial z}\right)^2\right]  &+&  \frac{\partial \Phi}{\partial t}+ g\eta  =0 \text{, on } z = \eta,
    \\
    \hspace{-0.5cm}
    \label{eq:system1.4} \frac{\partial \Phi}{\partial x}\frac{\partial h}{\partial x}+\frac{\partial \Phi}{\partial z} &=& 0 \text{, on } z = -h(x) \quad . 
\end{eqnarray} 
\citet{Zakharov1968a} and \citet{Craig1993} have shown the FNPF problem can be expressed using only the variables on the free surface, i.e., $\eta(x,t)$ and $\tilde{\Phi}(x,t)\equiv \Phi(x,z=\eta,t)$, which is also known as the Zakharov formulation:
\begin{align}
    \hspace{-0.2cm}
    \frac{\partial \eta}{\partial t} & =-\frac{\partial \tilde\Phi}{\partial x}\frac{\partial \eta}{\partial x}+\tilde w\left[1+\left(\frac{\partial \eta}{\partial x}\right)^2\right] \text{, on } z = \eta(x,t), \label{eq:Zakharov.1} \\
    \hspace{-0.2cm}
    \frac{\partial \tilde\phi}{\partial t} & =-g\eta-\frac{1}{2}\left(\frac{\partial \tilde\Phi}{\partial x}\right)^2+\frac{1}{2}\tilde w^2\left[1+\left(\frac{\partial \eta}{\partial x}\right)^2\right] \text{, on } z = \eta, \label{eq:Zakharov.2} 
\end{align}
where $\tilde w(x,t) \equiv \partial\Phi/\partial{z}(x,z=\eta(x,t),t)$ denotes the vertical velocity of the fluid particles on the free surface. To evaluate the temporal evolution of $\eta$ and $\tilde\Phi$, one needs to solve $\tilde w$ from the given free surface variables $(\eta, \tilde {\Phi})$. This is the core of the Zakharov formulation and is known as the Dirichlet--to--Neumann (DtN) problem. Various approaches have been proposed to solve the DtN problem \citep[see e.g.][and references therein]{Dommermuth2000, Madsen2006, Bingham2009, Papoutsellis2018}.

Recently, the exact FNPF problem has been resolved by using a spectral approach in the vertical direction and a high-order finite difference method in the horizontal direction, with a code called Whispers3D (denoted W3D hereafter). This model has been described and applied in various scenarios by \citet{Yates2015, Raoult2016, Simon2019, Zhang2021, Zhang2022}, showing high fidelity in modelling highly nonlinear non-overturning waves for arbitrary bottom profiles. 
In this model, a change of the vertical coordinate from $z$ to $s$ is adopted such that the physical domain $(x,z)$ with varying free surface and bottom boundaries $[-h(x), \eta(x,t)]$ can be transformed into a new domain $(x,s)$ with fixed upper and lower boundaries $[-1, 1]$:
\begin{equation}
  \label{eq:wishpers_s}
  s(x,z,t)=\frac{2z+h(x)-\eta(x,t)}{h(x)+\eta(x,t)}.
\end{equation}

Then, the model equations are reformulated in the $(x,s)$ domain and the vertical profile of the velocity potential is approximated using a basis of orthogonal Chebyshev polynomials of the first kind $T_n(s)$, truncated at a tunable maximum order $N_T$:
\begin{equation}
  \label{eq:spectral_approx}
  \Phi(x,s,t) \approx \sum_{n=0}^{N_T}a_n(x,t) T_n(s).
\end{equation}
The water wave problem is solved once the $N_T+1$ unknown coefficients $a_n(x,t)$ are determined at each abscissa $x$ from the DtN problem. The detailed reformulation of the governing equations in $(x,s)$ domain, the Chebyshev-tau method used to solve the DtN problem, and the numerical algorithm adopted in this model have been reported in the above-cited references, and are not duplicated here.

\begin{figure*}
\centering
    \includegraphics[width=0.9\textwidth]{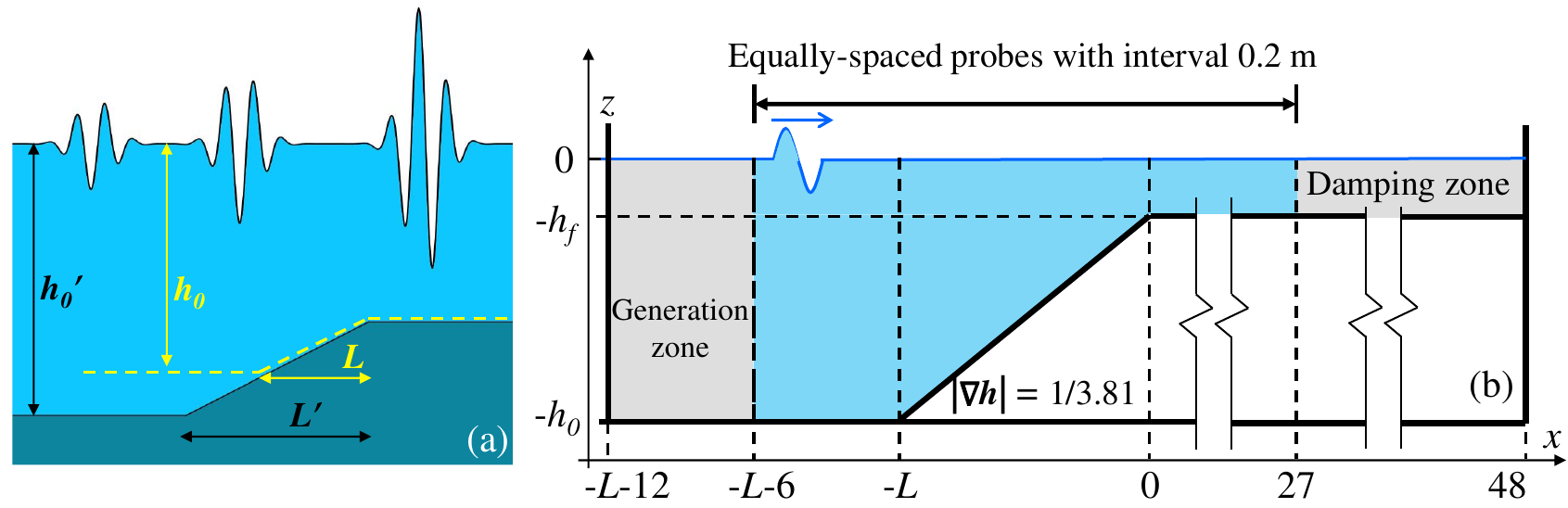}
    \caption{Sketches of (a) the necessary change in the bottom profile for assessing the role played by the shoaling length effect and (b) the numerical wave flume (NWF) (not to scale).}
\label{Fig:Bathy}
\end{figure*}
The parameter $N_T$ plays a crucial role in balancing the accuracy and efficiency of W3D, and should be adapted for different scenarios to achieve optimal model performance. To illustrate the role played by $N_T$, we present the comparison of the phase velocity $C(\mu \equiv kh,N_T)$ of sinusoidal waves in uniform water depth $h$ predicted by the linear wave version of W3D model with the analytical phase velocity $C_{Airy}(\mu)$ of Airy linear wave theory. As shown by \citet{Benoit2017}, the linear W3D model yields an analytical solution of $C(\mu,N_T)$: 
\begin{equation}
\label{eq:disper_NT}
\frac{C}{\sqrt{gh}}  = \sqrt{\frac{1 + \sum_{n=1}^{N_T-2}{A_n\mu^{2n}}}{1+\sum_{n=1}^{N_T-1}{B_n\mu^{2n}}}},
\end{equation}
where the $A_n$ and $B_n$ can be computed analytically \citep{Benoit2017}. The reference phase velocity from Airy theory reads:
\begin{equation}
    \frac{C_{Airy}}{\sqrt{gh}}=\sqrt{\frac{\tanh\mu}{\mu}}.
\end{equation}

Figure~\ref{Fig:Disper} shows that the relative error between $C$ and $C_{Airy}$ for linear waves reduces as $N_T$ increases, and it is of low level for deep-water waves (for instance, it is below $10^{-5}$ with $N_T \ge 7$ when $\mu=\pi$). This relative error remains acceptable even for extremely deep water when $\mu$ reaches $100$: it does not exceed $2.5\%$ with $N_T \ge 10$ and decreases down to about $1\%$ with $N_T=15$. It could be further improved at the expense of computational resources (by increasing $N_T$). Figure~\ref{Fig:Disper} also provides guidance for the selection of $N_T$ in different cases. The present study on the shoaling length parameter effects benefits from the W3D model with adjustable accuracy/efficiency, as the relative water depth varies significantly among the different simulation cases.

\section{Numerical simulations and results analysis \label{sec:analysis}}
\subsection{Configurations of the simulation cases\label{ssec:simulation_setup}} 

\begin{table*}
  \begin{center}
\def~{\hphantom{0}}
  \begin{tabular}{cccccccccccc}
  \toprule
   \multirow{2}{*}{Case} & \multirow{2}{*}{$h_0$ [m]} & \multirow{2}{*}{$L$ [m]} & \multirow{2}{*}{$\ell_p$} & \multirow{2}{*}{$\mu(4f_p,h_0)$} & \multicolumn{3}{c}{Deeper region} & \multicolumn{3}{c}{Shallower region}  & \multirow{2}{*}{$\mu_0/\mu_f$} \\
   \cmidrule(r){6-8}  \cmidrule(r){9-11}
    &  &  &  &  & $H_{s,0}$ [m] & {$\varepsilon_0$} & {$\mu_0$} & $H_{s,f}$ [m] & {$\varepsilon_f$} & {$\mu_f$} & \\
    \midrule
   1 & 0.13625 &  0.1 &  0.09 &  7.25 & 0.0190 & 0.0459 & {0.73} &  0.0193 & 0.0512 & {0.64} &  1.13\\
   2 & 0.16250 &  0.2 &  0.19 &  8.65 & 0.0191 & 0.0428 & {0.81} &  0.0197 & 0.0522 & {0.64} &  1.25\\
   3 & 0.21500 &  0.4 &  0.37 & 11.44 & 0.0188 & 0.0380 & {0.96} &  0.0198 & 0.0524 & {0.64} &  1.49\\
   4 & 0.26750 &  0.6 &  0.56 & 14.23 & 0.0186 & 0.0349 & {1.11} &  0.0197 & 0.0522 & {0.64} &  1.72\\
   5 & 0.32000 &  0.8 &  0.75 & 17.03 & 0.0186 & 0.0330 & {1.25} &  0.0196 & 0.0520 & {0.64} &  1.95\\
   6 & 0.42500 &  1.2 &  1.12 & 22.62 & 0.0191 & 0.0315 & {1.55} &  0.0199 & 0.0526 & {0.64} &  2.40\\
   7 & 0.53000 &  1.6 &  1.49 & 28.20 & 0.0193 & 0.0306 & {1.85} &  0.0198 & 0.0525 & {0.64} &  2.87\\
   8 & 0.74000 &  2.4 &  2.24 & 39.38 & 0.0198 & 0.0302 & {2.50} &  0.0199 & 0.0526 & {0.64} &  3.87\\
   9 & 0.95000 &  3.2 &  2.98 & 50.55 & 0.0202 & 0.0305 & {3.17} &  0.0199 & 0.0529 & {0.64} &  4.92\\
   10& 1.79000 &  6.4 &  5.97 & 95.25 & 0.0202 & 0.0304 & {5.95} &  0.0198 & 0.0525 & {0.64} &  9.24\\
   11& 3.47000 & 12.8 & 11.93 &184.65 & 0.0201 & 0.0303 &{11.54} &  0.0196 & 0.0520 & {0.64} & 17.91\\
  \bottomrule
  \end{tabular}
  \caption{\centering{Summary of the key parameters for the simulations. The incident sea-states are described by a JONSWAP spectrum of the same peak period $T_p=1.1$~s, and peak enhancement factor $\gamma = 3.3$. The slope length changes from Case 1 to 11, yet the slope gradient is kept constant $|\nabla{h}| = 0.2625$ (approximately 1:3.81). The incident $H_{s,0}$ is tuned in each case to keep $H_{s,f}$ more or less the same. The steepness measure can be converted to other common definitions through $\varepsilon = (\sqrt{2}/\pi)k_pH_s$. Note that case 7 here shares the same wave and bottom configurations as case 4 in \citet{Zhang2022}.}}
  \label{tab:cases3}
  \end{center}
\end{table*}
The key to choosing the wave parameters of the incident wave train and the bathymetry setup lies in two aspects. On one hand, the incident wave train parameters are manipulated to keep the steepness $\varepsilon$ and relative water depth $\mu$ after the shoal nearly unchanged among the various cases, such that the sea-states after the shoal differ only because of different degrees of non-equilibrium dynamics induced by different lengths of the shoal. On the other hand, the slope magnitude and shoaling length are disentangled by keeping $|\nabla{h}|$ constant and varying only the slope length $L$. Furthermore, we choose a bottom profile with an upslope only but no downslope to avoid any confusion due to the disturbances of the reflected wave energy during the de-shoaling process \citep{Zhang2021}.

Inspired by run 3 of the experiments of \cite{Trulsen2020} and the simulations of \cite{Zhang2022}, we keep the water depth after the shoal $h_f=0.11$~m and incident peak period $T_p=1.1$~s as constants in all cases, thus the relative water depth in the shallower region $\mu_f$ is also a constant. In theory, $\ell$ varies over the slope and should be computed locally in space. Hereafter, we use its value at the end of the shoal, $\ell_{p} \equiv L/\lambda_{p,f}$, to better show the correspondence between the enhancement of kurtosis and the shoaling length parameter. As a result, the change in bathymetry now boils down to a longer shoal and a deeper water depth $h_0$ prior to the shoal, as illustrated in figure~\ref{Fig:Bathy}(a).
Suppose the slope length $L$ is extended by a factor of $n$ to $L' = nL$, resulting in a new deeper region depth $h_0'$. Since $|\nabla h|$ is kept the same in both cases, i.e. $(h_0'-h_f)/L' = (h_0-h_f)/L$, the pre-shoal depth is scaled correspondingly:
\begin{equation}
\frac{h_0'}{h_0}=n \left[ 1 + \left(\frac{1}{n}-1\right)\frac{h_f}{h_0} \right] \quad , \quad n \, \in \, \mathbb{R}^{\ast}_{+} \,\, .   
\end{equation}
As the change of slope length will lead to different shoaling coefficient (defined as $C_{\textrm{shoal}} = H_{s,f}/H_{s,0}$), the incident $H_{s,0}$ is tuned in each case, such that $\varepsilon_f$ atop the shoal is kept the same regardless of the length of the shoal.

We consider long-crested irregular wave trains described by a JONSWAP spectrum: 
\begin{equation}
\label{eq:jonswap}
  S(f)=\frac{\alpha_J{g}^2}{\left(2\pi\right)^4}\frac{1}{f^5}\exp{\left[-\frac{5}{4}\left(\frac{f_p}{f}\right)^4\right]}\gamma^{\exp{\left[-\frac{\left(f-f_p\right)^2}{2\left(\sigma_J f_p\right)^2}\right]}},
\end{equation}
where $\alpha_J$ denotes the adjustment factor for $H_s$, $\sigma_J$ the spectral asymmetry parameter ($\sigma_J=0.07$ for $f \leq f_p$ and $\sigma_J=0.09$ for $f>f_p$), and $\gamma$ the peak enhancement factor. In this study, the spectral peak period $T_p=1/f_p=1.1$~s and peak enhancement factor $\gamma=3.3$ of the incident wave field are the same for all 11 cases. In each case, the incident wave train lasts for 5060 s, thus $4600T_p$. Table~\ref{tab:cases3} summarizes the configurations of the incident wave fields and bathymetry information in 11 cases, as well as the key wave parameters in both deeper and shallower regions (by averaging the simulation results over the corresponding areas). These cases are considered representative for investigating the shoaling length effect, as they cover a relatively wide range of shoaling length parameter, $\ell_p \in [0.1, 11.9]$. Meanwhile, the relative water depth difference before and after the shoal also varies significantly, $\mu_0/\mu_f \in [1.1, 17.9]$.

The incident wave train is constructed by linearly superimposing the harmonic components of the prescribed spectrum, and imposed at the wave-making boundary of the numerical wave flume (NWF). The low-energy components of the incident spectrum are ignored, keeping the non-trivial ones in the range $[f_{min}, f_{\textrm{max}}]=[0.5f_p,4f_p]$, as justified in the next sub-section. The sketch of the NWF is provided in figure~\ref{Fig:Bathy}(b): the water depth $h_0$ is constant in the flat area prior to the shoal (6~m in length, roughly $3\lambda_{p,0}$), then it decreases to $h_f = 0.11$~m due to the presence of an up-slope (with a constant slope magnitude $|\nabla h| =0.2625$ and length $L$), and remains constant in the flat area (27~m in length, roughly $25\lambda_{p,f}$) after the shoal. Such a long after-shoal flat area allows the out-of-equilibrium sea-state, induced by the depth change, to re-establish a new equilibrium state, based on the work in \citet{Zhang2022}. In addition, the computation domain comprises a generation relaxation zone of 6~m in length ($\approx 3\lambda_{p,0}$) on the left end of the NWF and an absorption relaxation zone of 21~m in length ($\approx 20\lambda_{p,f}$) on the right end. As detailed in appendix~\ref{sec:append_LF}, such a length of the damping zone is needed to keep negligible reflection of all wave components, including the low-frequency ones, in spite of the higher cost of computation effort.

\subsection{Result convergence and statistical variability \label{ssec:convergence}} 

Case 7 in the present study shares the same wave and bottom conditions as case 4 in \citet{Zhang2022}, and other cases are also very similar to it (see Table~\ref{tab:cases3}).
Therefore, it is logical to adopt the same spatial grid interval and time step, d$x = 0.01$~m and d$t=0.01$~s, which result in the Courant-Friedrichs-Lewy number $\text{CFL}_i = C_{p,i} \text{d}{t}/ \text{d}{x} = (\lambda_{p,i}/ \text{d}{x})/(T_p/ \text{d}{t})$, with $i=0$ or $f$, to be $\text{CFL}_0 =1.64$ prior to the shoal and $\text{CFL}_f=0.97$ atop the shoal. 

\begin{figure*}
\centering
    \includegraphics[width=0.8\textwidth]{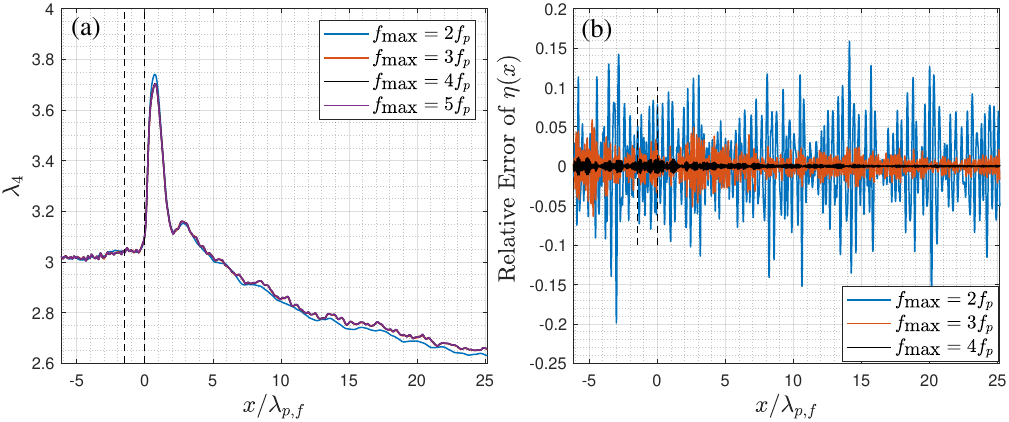}
    \caption{Panel (a): spatial evolution of kurtosis $\lambda_4$ for case 7 considering four increasing cut-off frequencies $f_{\textrm{max}}$ of the incident JONSWAP wave spectrum; panel (b): relative error of $\eta(x)$ at final time $t = 5060$~s for the simulations with $f_{\textrm{max}} = 2 f_p, 3 f_p$ and $4 f_p$, with respect to the simulation with $f_{\textrm{max}} = 5 f_p$ used as reference.}
\label{Fig:Converge_cutoff}
\end{figure*}
The investigation of the shoaling length effect poses high requirements on the model accuracy. The increase of slope length results in larger $\mu_0$, and the demands on model accuracy increase sharply to describe the wave evolution in the deeper flat region. This is because, for irregular waves, the model should be reasonably accurate up to the cut-off frequency $f_{\textrm{max}}$ of the input spectrum, which is much more challenging than considering model accuracy up to the spectral peak frequency $f_p$. Table~\ref{tab:cases3} lists the relative water depth $\mu(4f_p,h_0)$ for waves with $f=4f_p$ in depth $h_0$. In case 11 for instance, the relative water depth corresponding to $f_{\textrm{max}}$ in the NWF is actually higher than 180 prior to the shoal. We know from figure~\ref{Fig:Disper} that the accuracy of W3D is sensitive to the relative water depth, and that larger $N_T$ is needed for deeper water. We therefore performed a series of convergence tests for determining the cut-off frequency of the input spectrum and the choice of $N_T$ value, both would influence the performance of W3D.

\begin{figure*}
\centering
\includegraphics[width=0.8\textwidth]{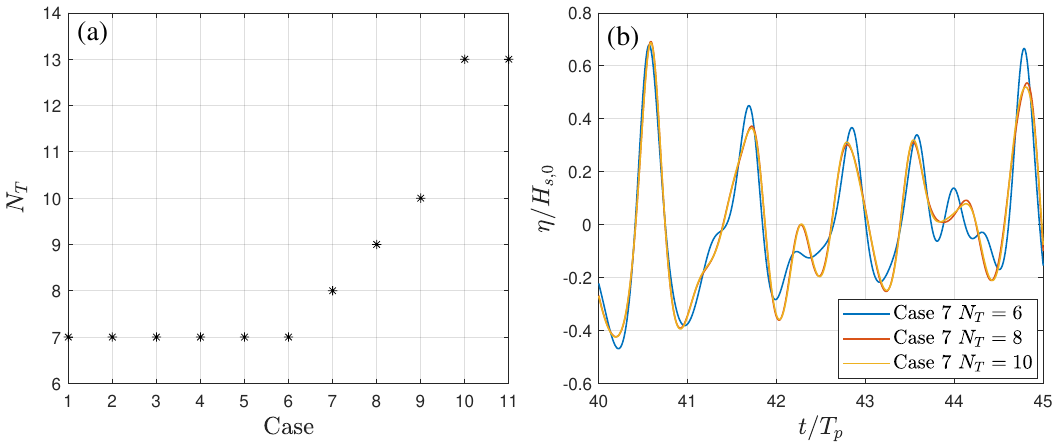}
    \caption{Panel (a): calibrated values of $N_T$ for cases 1 to 11; panel (b): comparison of snapshots of duration $5 T_p$ of the time series of FSE $\eta$ at 6~m after the shoal in case 7 simulated with $N_T=$ 6, 8, 10.}
\label{Fig:converge_NT}
\end{figure*}
We determined the extent of the spectral range $[0.5f_p,f_{\textrm{max}}]$ required to ensure accurate convergence of the simulations by running case 7 with $f_{\textrm{max}}$ taking $2f_p$, $3f_p$, $4f_p$ and $5f_p$ successively. Figure~\ref{Fig:Converge_cutoff}(a) shows the comparison of the spatial evolution of $\lambda_4$ with these four different cut-off frequencies. It is observed that the estimation of kurtosis is convergent for $f_{\textrm{max}} \ge 3f_p$. Figure~\ref{Fig:Converge_cutoff}(b) shows the relative error of $\eta(x)$ at $t=5060$~s with respect to the result with highest $f_{\textrm{max}}=5f_p$, and normalized by the incident significant wave height $H_{s,0}$. It is noticed that the convergence of FSE in the time domain is closely achieved for $f_{\textrm{max}}=4f_p$ with relative error within 1\%. Therefore, $f_{\textrm{max}}=4f_p$ is chosen for all subsequent simulations.
Furthermore, the evolution of $\lambda_4$ in the case with $f_{\textrm{max}}=2f_p$ differs only slightly from the well-converged results with $f_{\textrm{max}} \ge 4f_p$. This seems to be contradictory to the conclusion of \citet{Adcock2022}, which states that removing the high-frequency spectral tail would result in substantial enhancement of $\lambda_4$. In fact, this is because the mechanisms are different in the study of \citet{Adcock2022} and in the present one: in the former work, the spectral change manifests spontaneously under the flat bottom condition due to strong nonlinear wave-wave interaction. Whereas, in the latter, the spectral modulation is forced by the depth transition, and thus is less sensitive to the high-frequency tail of the incident spectrum but more to the rapid depth change.

All the cases listed in Table~\ref{tab:cases3} have been first tested for a short duration $50$~s (about $45 T_p$) to tune the $N_T$ values so that the balance between efficiency and precision can be achieved in each case. In figure~\ref{Fig:converge_NT}(a), the tuned values of $N_T$ are displayed for all 11 cases. In figure~\ref{Fig:converge_NT}(b), the calibration of $N_T$ in case 7 is provided as an example, it is observed that the time evolution of $\eta$ is identical for $N_T = 8$ and $10$, therefore in the final simulation of case 7 with long duration, $N_T=8$ is chosen. 

\begin{figure*}
\centering
\includegraphics[width=0.85\textwidth]{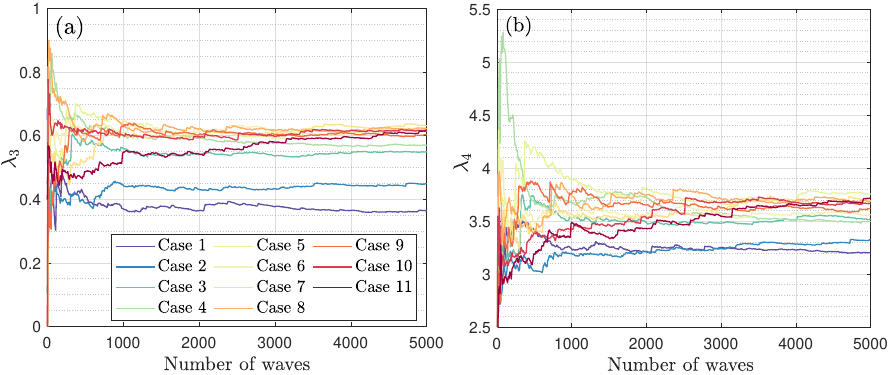}
    \caption{\centering{Convergence of (a) skewness and (b) kurtosis of FSE at $x=0.75$ m (equivalently $x/\lambda_{p,f} = 0.7$, where the skewness and kurtosis assume their maximum values) as a function of the number of waves in the simulated time series in cases 1 to 11.}}
\label{Fig:converge}
\end{figure*}

The skewness and kurtosis, as high-order moments, are subject to statistical variability. The duration of the record of $\eta$ plays an important role in obtaining (at least nearly) converged results of skewness and kurtosis. We checked that a time duration of 5060 s (equivalently, 4600$T_p$) in the simulations is sufficient for obtaining statistically converged estimates of skewness and kurtosis. For such a purpose, the convergence of skewness and kurtosis is analysed with the time series computed at the location $x=0.75$~m where they assume their maximum values and are subject to the strongest variability. The time series of $\eta$ is separated into a series of individual waves through the zero-down-crossing method. Then, by evaluating the skewness and kurtosis of a section of the entire time series that contains the first $N$ waves, the skewness and kurtosis are functions of the number of waves in the time series. In figure~\ref{Fig:converge}, $\lambda_3$ and $\lambda_4$ computed from different numbers of waves in the time series are shown, it is noted that the estimates become nearly stable after including 4000 random waves, thus the selected simulation duration is long enough.

\subsection{Shoaling length effects on the spectral evolution along the NWF \label{ssec:spectral_analysis}} 
\begin{figure*}
\centering
    \includegraphics[width=0.75\textwidth]{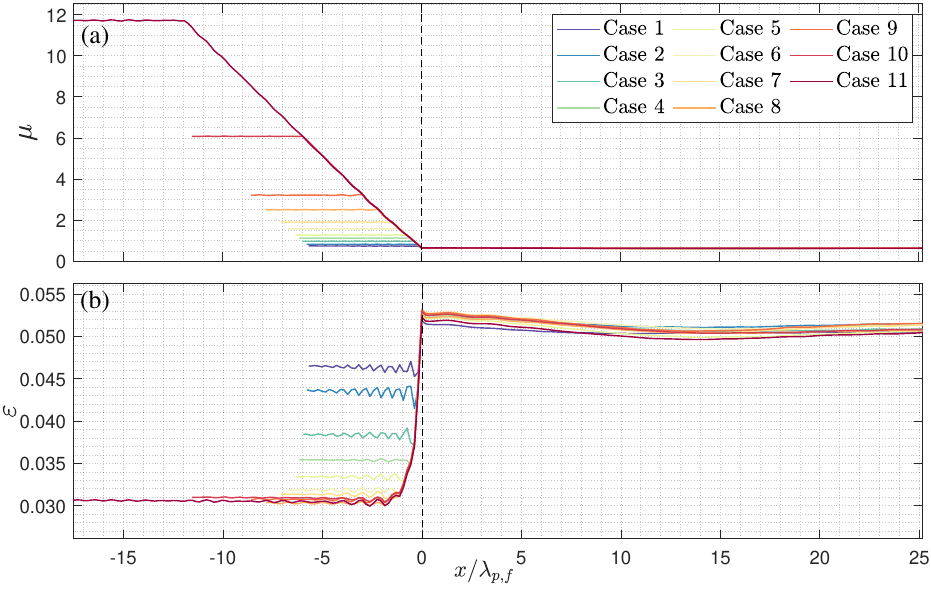}
    \caption{\centering{Spatial evolution along the NWF of non-dimensional wave parameters for cases 1 to 11: (a) relative water depth $\mu$ and (b) wave steepness $\varepsilon$. The vertical dash line at $x=0$ indicates the starting location of the shallower flat region. }}
\label{Fig:spec_para}
\end{figure*}

The spatial evolution trends of two key wave parameters, the relative water depth $\mu$ and the wave steepness $\varepsilon$, are displayed in figure~\ref{Fig:spec_para}. They are evaluated locally, based on the spectral peak frequency and significant wave height extracted from spectral analysis of the time series of FSE. This figure further validates the methodology for determining the wave parameters of the input wave field, which ensures that $\mu_f$ and $\varepsilon_f$ after the shoal are basically the same whatever $\ell$. Before the shoal, $\mu$ varies significantly due to the change of depth $h_0$ among the cases, and $\varepsilon$ also differs in order to compensate for different shoaling effects on $H_{s,f}$ among the cases.
\begin{figure*} 
\centering
    \includegraphics[width=0.9\textwidth]{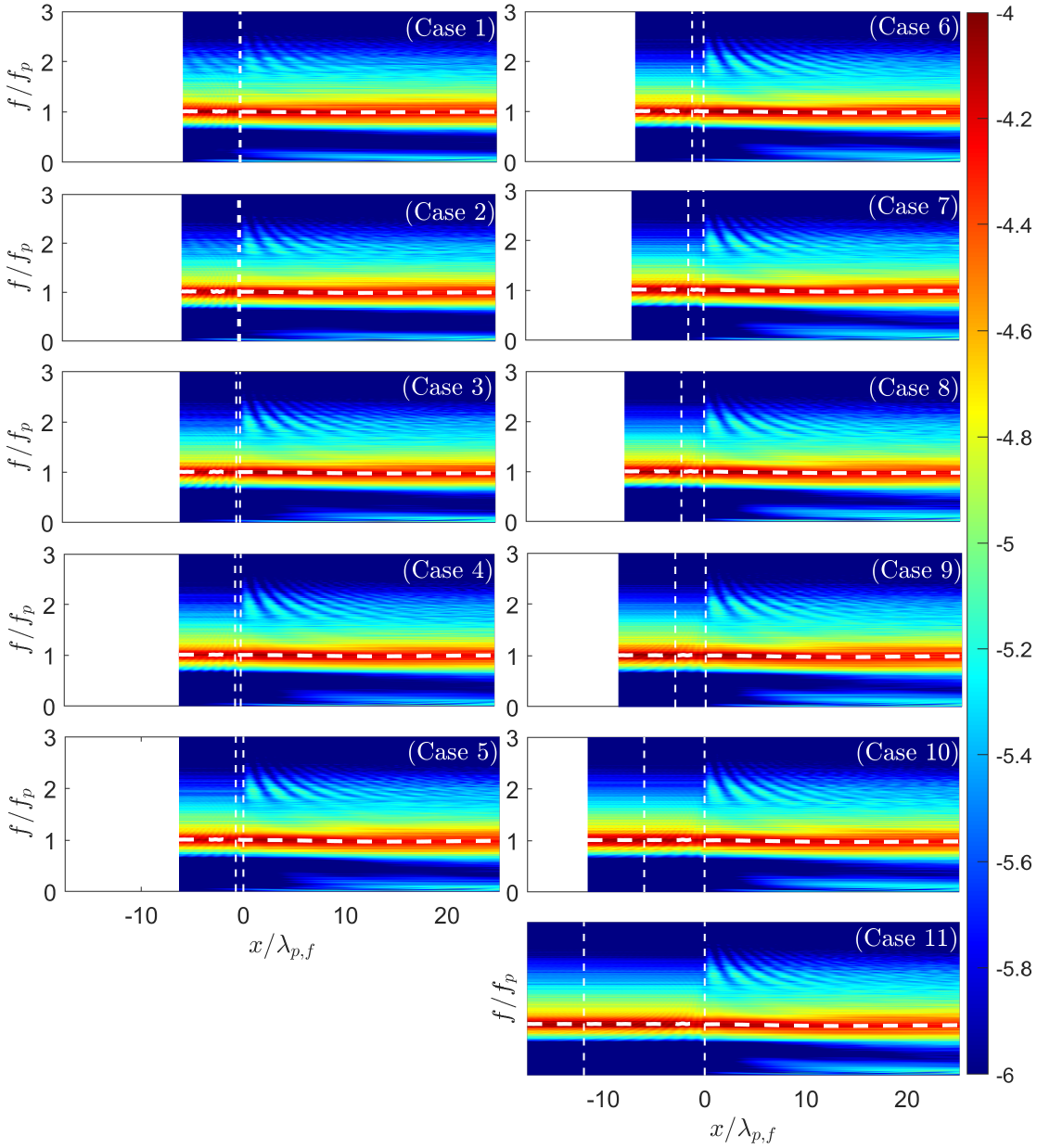}
    \caption{\centering{Spatial evolution along the NWF of the wave spectrum for cases 1 to 11, displayed in panels (a) to (k) respectively. The color scale depicts the value of log$_{10}(S(f,x))$ with spectrum $S(f,x)$ in m$^2$/Hz. The vertical white dash lines represent the extent of the plane slope. The horizontal dash line represents the local peak frequency of the wave spectrum.}}
\label{Fig:spec_space}
\end{figure*}
\begin{figure*} 
\centering
    \includegraphics[width=0.6\textwidth]{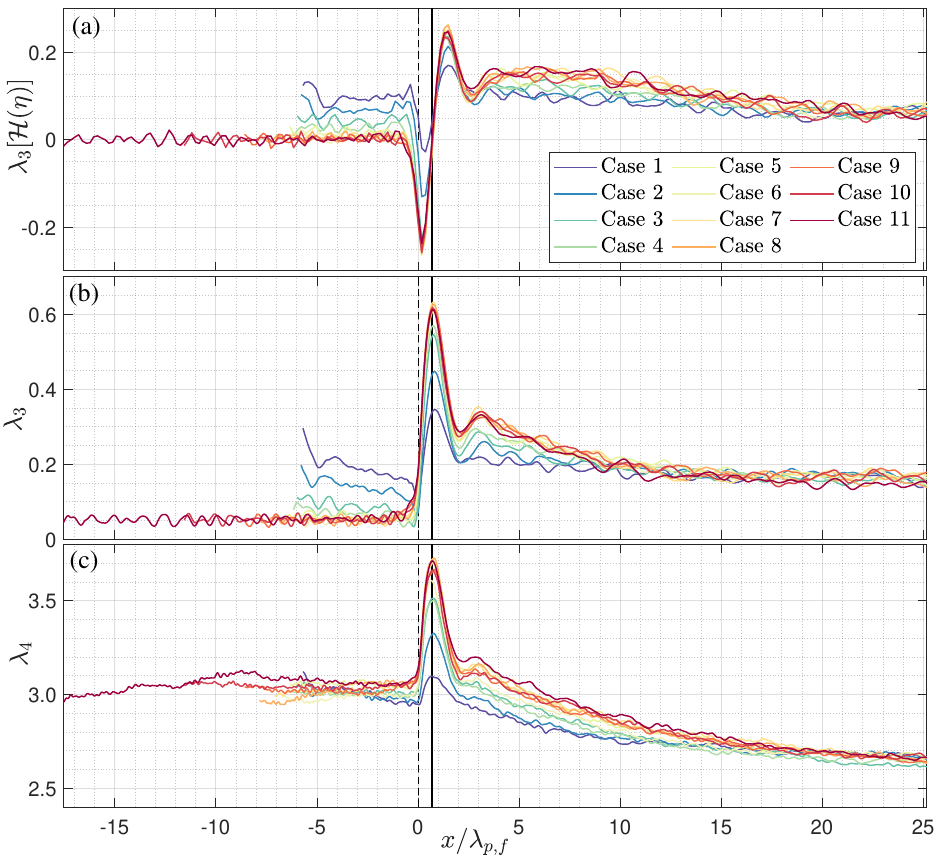}
    \caption{Spatial evolution of (a) asymmetry parameter $\lambda_3[\mathcal{H}(\eta)]$, (b) skewness $\lambda_3(\eta)$ and (c) kurtosis $\lambda_4(\eta)$, in cases 1 to 11 along the NWF. The vertical solid line at $x=0.7\lambda_{p,f}$ indicates the position where skewness and kurtosis achieve their maxima.}
\label{Fig:three_stat}
\end{figure*}
The FSE time series is saved every $0.2$~m (0.2$\lambda_{p,f}$ approximately) in the NWF, resulting in a relatively fine resolution of the spectral evolution in space, as is displayed in figure~\ref{Fig:spec_space}. It is noted that, in all cases, the spectral evolution after the shoal is almost the same: a beating pattern manifests for the second harmonics shortly after the shoal then gradually vanishes as waves propagate further, eventually, a broadened spectrum is established. It indicates that the shoaling length effect is insignificant for wave spectral evolution as waves pass over a steep shoal. In cases 1 and 2, the spectral evolution is slightly different from the other cases, i.e., the beating pattern appears not only in the range after the slope but also in the flat area before it. This is because in both cases, the relative water depths $\mu_0$ before the shoal are the lowest two among others, meanwhile, the steepness values $\varepsilon_0$ are higher than other cases, resulting in a higher relative importance of wave nonlinearity over dispersion.

\subsection{Shoaling length effects on the spatial evolution of statistical parameters \label{ssec:slope_effects}} 

Here, the spatial evolution of the asymmetry parameter, skewness and kurtosis is investigated, and the influences of $\ell$ parameter on the evolution of these key statistical parameters are discussed. The asymmetry parameter is a quantity analogous to skewness, which measures the wave profile asymmetry in the horizontal direction. It is computed as the skewness of the Hilbert transform of the FSE, $\lambda_3[\mathcal{H}(\eta)]$, with $\mathcal{H}$ denoting the Hilbert transform operator. The asymmetry parameter assumes negative values when the wave front is steeper than its rear face, for instance, when a wave passes over a shoal. The results of the three statistical parameters are displayed in the subplots (a--c) of figure~\ref{Fig:three_stat}.

In figure~\ref{Fig:three_stat}(a), the evolution of the asymmetry parameter is displayed for all cases. The horizontal wave profile first leans forward due to shoaling (with $\lambda_3[\mathcal{H}(\eta)]$ approaching its negative extreme value) in the short area near the end of the slope, and then the wave profile develops reversely (with $\lambda_3[\mathcal{H}(\eta)]$ returning to 0 around $x=0.7\lambda_{p,f}=0.75$~m on its way to the positive maximum), and becomes backwards leaning due to significant non-equilibrium wave evolution. Eventually, $\lambda_3[\mathcal{H}(\eta)]$ approaches a positive constant in the equilibrium state that is nearly established close to the end of the NWF. All cases 5 to 11 follow this trend, with their positive and negative global extreme values and the equilibrium values being almost the same and achieved at the same location. The picture is a bit different in cases 1 to 4, where the asymmetry parameter is higher than 0 initially due to the higher relative importance of nonlinearity as illustrated previously.

The spatial evolution of skewness $\lambda_3(\eta)$ in all cases is shown in figure~\ref{Fig:three_stat}(b). It has a finer resolution compared to figure~\ref{Fig:three_stat}(a), because W3D allows the computation of skewness and kurtosis at every grid point (with d$x=0.01$~m) during the simulation. As for the asymmetry parameter, the evolution trends are very similar in all cases, with their maximum values atop the shoal converging at the same location (convergent maximum value $\lambda_3 \approx 0.6$). Furthermore, the equilibrium values of $\lambda_3$ in the far field atop the shoal are also at the same level, as the sea-states in all cases are of nearly the same levels of nonlinearity and dispersion. In cases 1 to 4, the skewness in the deeper flat region before the slope is higher than that in other cases, this is again related to the higher relative importance of wave nonlinearity.

In figure~\ref{Fig:three_stat}(c), the spatial evolution of kurtosis in all cases is shown. In all cases, $\lambda_4$ remains around 3 before the end of the shoal, and after the shoal, $\lambda_4$ is significantly enhanced to its local maximum at $x=0.7\lambda_{p,f}=0.75$~m, indicating a local increase of the rogue wave occurrence probability. Then, $\lambda_4$ declines to a level around 2.6, which indicates that the probability of rogue waves in the new equilibrium state is even lower than that in a Gaussian sea-state. We notice that $\lambda_4$ may not be convergent yet at $x=25\lambda_{p,f}$. This is understandable because as a fourth-order moment, kurtosis would require a longer distance to be convergent than the third-order ones. As the trend of kurtosis is already clear in figure~\ref{Fig:three_stat}(c), we decided not to extend the length of NWF. Based on the evolution of these three statistical parameters shown in figure~\ref{Fig:three_stat}, we consider that the short-scale non-equilibrium wave evolution stage happens in the range $[0,5]\lambda_{p,f}$ and the long-scale in $[5,25]\lambda_{p,f}$ in the NWF. The spatial extent of these two scales is independent of $\ell_p$.

Now we focus on the local extremes of these statistical moments. From figure~\ref{Fig:three_stat}(a), it is noticed that the local extreme values of $\lambda_3[\mathcal{H}(\eta)]$ after the shoal in cases 1 and 2 are of lower modulus than other cases. From figures \ref{Fig:three_stat}(b) and (c), we see that in cases 1 to 4, $\lambda_3$ and $\lambda_4$ are of the lower maximum values shortly after the shoal. This indicates that the depth changes in cases 1 to 4 are in a transitional regime in which the water body switches from a homogeneous media (constant depth condition) to an inhomogeneous media (strongly varying depth condition). When the change of depth is large enough, in the simulated cases this means $\ell_p \ge 0.5$, the wave propagation after the shoal is dominated by the non-equilibrium dynamics induced by the media inhomogeneity. For $\ell_p \ge 0.5$, the maximum of skewness and kurtosis, and thus the probability of rogue waves, is no longer dependent upon the shoaling length parameter $\ell_p$.

\begin{figure*}
\centering
    \includegraphics[width=0.65\textwidth]{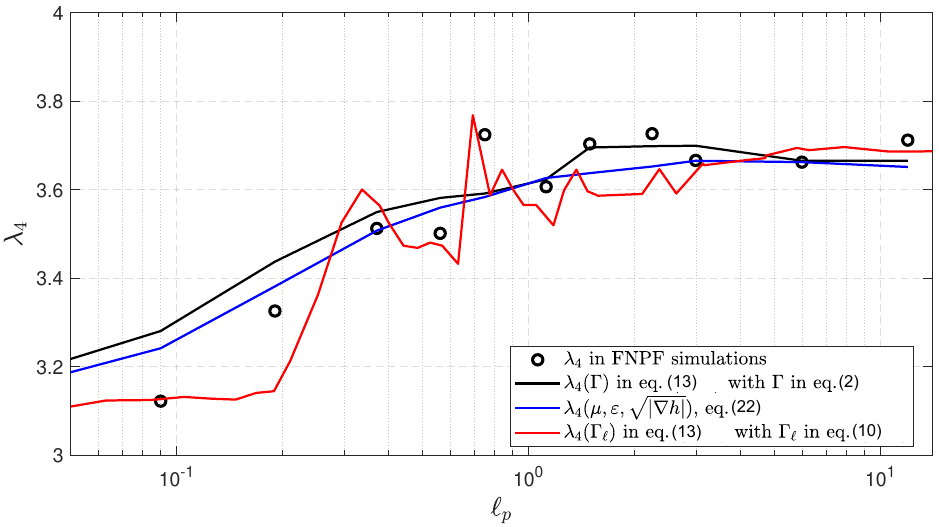}
    \caption{Comparison of the maximum values of $\lambda_4$ atop the shoal between simulated results and theoretical predictions, displayed as functions of the shoaling length parameter $\ell_p$. The simulation results are marked as hollow circles, and the kurtosis predictions are computed in three ways (as indicated in the legend box)
    }
\label{Fig:SK_L}
\end{figure*}

In figure~\ref{Fig:SK_L}, the maximum value of $\lambda_4$ in the short scale is displayed as a function of $\ell_p$. The empirical maximum kurtosis is extracted from the simulations in the range $x\in[0,5]\lambda_{p,f}$. The theoretical predictions of $\lambda_4$ can be obtained in three ways: (1) with $\lambda_4$ in eq.~(\ref{eq:Mori3X}) and $\Gamma$ in eq.~(\ref{eq:Rayexc}), originally put forward in \citet{Mendes2023} without dependence on $\ell$; (2) with $\hat{\lambda}_4$ in eq.~(\ref{eq:Mori3X}) and newly developed $\Gamma_{\ell}$ in eq.~(\ref{eq:gammaL}), with the shoaling length effect included in $\Gamma_{\ell}$; and (3) with the simplified expression of kurtosis, eq.~(\ref{eq:Mori3Xapprox}) as an approximation to option (2). As a remark, the $\ell$ in eq.~(\ref{eq:gammaL}) is computed from a zero-crossing wavelength, whereas the simulations have a peak wavelength counterpart. This can be amended by using empirical relations between the two measures, and following \citet{Figueras2010} for the relation between peak and mean periods, as well as \citet{Mendes2021} for the relation between mean and zero-crossing period, we may approximate $T_{p}^{2} \approx 2 T_{z}^{2}$ (with $T_z$ denoting the zero-crossing mean wave period) such that $\ell \sim 2 \ell_{p}$. In figure~\ref{Fig:SK_L}, the agreement between the empirical values of kurtosis and the theoretical predictions with either the original formulation eq.~(\ref{eq:Mori3X}) without $\ell$ dependence or the approximated formulation eq.~(\ref{eq:Mori3Xapprox}) is good for all cases, despite that the empirical kurtosis closely scatters around the theoretical results. The theoretical prediction of $\lambda_4$ with eqs.~(\ref{eq:Mori3X}) and (\ref{eq:gammaL}) shows evident oscillation for $\ell_p <0.5$, this is related to the oscillatory feature of $\Gamma_{\ell}$ as illustrated in figure~\ref{Fig:ell}(a). As has been explained, the oscillations of the theoretically calculated $\Gamma_\ell$ arise from the local definition of wave energy over a range shorter than the wavelength in eq.~(\ref{eq:energyX0}).
Additionally, it is worth mentioning that the theoretical formulations are able to predict the maximum $\lambda_4$ even in cases 1 to 4, which are considered to be transitional cases from homogeneous to non-homogeneous conditions. We conclude that the shoaling length parameter plays an insignificant role in the maximum kurtosis atop the shoal at least for $\ell_p>0.5$ and that the approximated expression eq.~(\ref{eq:Mori3Xapprox}) provides rather accurate predictions of maximum kurtosis free from the spurious oscillation for all $\ell_p$.

The non-homogeneous second-order theory predicts the maximum kurtosis corrected by the shoaling length parameter, eq.~(\ref{eq:Mori3Xapprox}), is almost the same as the prediction without $\ell$ dependence via eqs.~(\ref{eq:Mori3X}) and (\ref{eq:Rayexc}). As a result, we conclude that the contribution of the shoaling length parameter $\ell$ to the non-equilibrium statistics atop the shoal is small even in cases 1 to 4 with $\ell_p<0.5$. The transitional regime is attributed to the insufficient depth prior to the shoal, such that the non-equilibrium dynamics is weak. This complements the results from the numerical simulations, which by themselves cannot distinguish this interpretation from a potential saturation of the effect of the shoaling length for $\ell_p \ge 0.5$.

As a remark, the small but tangible drop of the red curve in comparison with the black/blue curves in figure~\ref{Fig:SK_L} for $\ell_p \le 0.3$ could be caused by the faster transmission of the entire wave energy to such a short scale, thereby decreasing $\Gamma$. This is acceptable, because on the one hand, from a cost-benefit perspective, the model described by the blue curve is still more practical as it is not computationally burdensome. On the other hand, from a practical point of view, non-equilibrium dynamics associated with such a small $\ell_p$ is not the main physics one should care about.
For the non-equilibrium wave evolution induced by depth transitions, there are three stages which are characterized by different features: (1) the near-equilibrium evolution stage before the end of the slope (the manifestation of non-equilibrium dynamics often lags behind the depth change, this feature is known as "latency", see \citet{Adcock2020} for instance); (2) the short-scale non-equilibrium evolution stage close to the end of the up-slope; and (3) the long-scale non-equilibrium evolution stage as waves propagate further before reaching a new equilibrium state. Both our FNPF simulation results and non-homogeneous second-order theory predictions have shown that the shoaling length parameter has no effect on the spatial extent of short- and long-scale non-equilibrium wave evolution, nor on the locations where global extreme values of the statistical parameters are achieved.

\section{Conclusion \label{sec:conclusion}}

In this work, we have shown both theoretically and numerically that the maximum enhancement of the kurtosis of FSE atop the shoal and therefore of the rogue wave probability are driven by the magnitude of the bottom slope magnitude rather than by the shoaling length.  
We provided an explicit definition of the abruptness of a shoal by introducing a shoaling length parameter $\ell$ as the ratio between the slope length and the characteristic wavelength (denoted $\ell_p$ when calculated with the wavelength at the peak frequency atop the shoal).
This allowed us to disentangle this shoaling length parameter from the slope magnitude by varying the pre-shoal depth while keeping the slope magnitude constant.

By building on the recent theoretical work in~\citet{Mendes2022}, we provided theoretical evidence that the rogue wave amplification is independent of the shoaling length, as long as the relative water depth, wave steepness atop the shoal, and slope magnitude remain constant. Then, we showed numerically that not only the rogue wave amplification (characterized by the peak values of the statistical moments) but also the location of these maxima, the vertical asymmetry in surface elevation, wave steepness, and spectral evolution are insensitive to the shoaling length. We observed that shorter shoaling lengths (typically $\ell_p \le 0.5$) lead to reduced increase of kurtosis over the shoal, since the depth prior to the shoal is insufficient to induce non-equilibrium dynamics. 

Finally, our study covers the range $0.1 \leqslant \ell_p \leqslant 12$, but equally applies beyond. Indeed, longer shoals correspond to even deeper pre-shoal conditions, where the waves do not interact with the bottom. The shoal therefore only starts to affect the propagation when the waves quit the deep-water region, thus limiting the effective length of the shoal. Conversely, very short $\ell$ implies negligible depth difference, hence, a vanishing shoal effect.

When designing the setup of the FNPF simulations, we have shown that (i) a cut-off at higher frequencies of the wave spectrum has a significantly smaller impact on wave statistics than expected for a flat bottom in deep water, and (ii) insufficient attenuation of low-frequency waves at the downstream boundary of the domain could have a notable influence on wave statistics and spectrum atop the shoal. Dedicated sensitivity studies and choices have been done to get rid of these two effects in the final set of simulations.

The outcomes of this work could also be useful for researchers investigating either experimentally or numerically the depth-induced non-equilibrium dynamics when designing the incident wave and bathymetry configurations. Also, the explicit and simple expression for the maximum kurtosis eq.~\eqref{eq:Mori3Xapprox} can be used for fast estimation of freak wave risk associated with a steep shoal in engineering practices. Further work would however be needed to assess the effect of $\ell$ on waves of larger steepness, especially reaching close to wave breaking.


\section*{Declaration of competing interest}
The authors declare that they have no known competing financial interests or personal relationships that could have appeared to influence the work reported in this paper.

\section*{Acknowledgments}
In this work, J.Z. was supported by the National Natural Science Foundation of China (Grant No. 52101301), and the China Postdoctoral Science Foundation (Grant No. 2023T160078, 2021M690523). S.M. and J.K. were supported by the Swiss National Science Foundation (Grant No. 200020-175697). 

\appendix

\section{Effect of shoaling length on energetics}\label{sec:append_abruptE}
In this section, we detail step-by-step the integration of variance and energy of FSE and assess its implications on the wave statistics through the measure of $\Gamma$ in eq.~(\ref{eq:Rayexc}). We plug the solutions of eqs.~(\ref{eq:seteq},\ref{eq:seteqx}) into eq.~(\ref{eq:energyX0}). As pointed out in \citet{Mendes2022b}, the physical variables $(a, k, \varepsilon)$ vary slowly as compared to the trigonometric functions as long as $H_\textrm{s} \ll h$
, hence, they are taken out of the integral. Thus, the integration is limited to the auxiliary integrals:
\begin{numcases}{}
\mathcal{I}_{1} \equiv \int_{0}^{L}  \cos^{2}{\phi} \, \frac{\textrm{d}x}{L} = \frac{1}{2} \left[ 1 + \frac{ \sin{(4\pi \ell)} }{4\pi \ell} \right] \, , \notag\\ \mathcal{I}_{2} \equiv \int_{0}^{L}  \cos^{2}{(2\phi)} \, \frac{\textrm{d}x}{L} = \frac{1}{2} \left[ 1 + \frac{ \sin{(8\pi \ell)} }{8\pi \ell} \right],
\label{eq:setint}
\\
\mathcal{I}_{12} \equiv \int_{0}^{L}  \cos{\phi} \, \cos{(2\phi)} \, \frac{\textrm{d}x}{L} = \frac{ 3\sin{(2\pi \ell)} +  \sin{(6\pi \ell)}}{12\pi \ell},  \notag\\
\mathcal{I}_{34} \equiv \int_{0}^{L}  \sin{\phi} \, \sin{(2\phi)} \, \frac{\textrm{d}x}{L} = \frac{  \sin^{3}{(2\pi \ell)}}{3\pi \ell}. \notag
\end{numcases}
The zero-order integrals containing $\cos{\phi} \cos{(2\phi)}$, such as $(\mathcal{I}_{12}, \mathcal{I}_{34})$, are not of leading order and are related to the mean water level correction. As in the case of assuming $\ell = 1$ in \citet{Mendes2022}, the squared trigonometric functions leading to $(\mathcal{I}_{1}, \mathcal{I}_{2})$ take the leading role in driving anomalous statistics because they still return non-zero values when the shoaling length integrals $(\mathcal{I}_{12}, \mathcal{I}_{34})$ vanish at half-integers $\ell = (2n+1)/2$, with $n \in \mathbb{N}$.

In the present configuration, the variance of FSE is approximated as a spatial average:
\begin{equation}
\label{eq:variance_app}
\langle \eta^{2} \rangle \approx \frac{1}{L} \int_{0}^{L}  \eta^{2}(x) \, \textrm{d}x.
\end{equation}

Given the expression of FSE at second-order \eqref{eq:seteq} and the identities of eq.~(\ref{eq:setint}), we obtain:
\begin{eqnarray}
\nonumber
\langle \eta^{2} \rangle &=&  a^{2} \left [  \mathcal{I}_{1} + 2\mathcal{I}_{12} \left(  \frac{\pi \varepsilon}{4} \right) \sqrt{ \Tilde{\chi}_{1} }  + \mathcal{I}_{2} \left(  \frac{\pi \varepsilon}{4} \right)^{2}  \Tilde{\chi}_{1} \right ]
\label{eq:energyVAR}
\\
\nonumber
&=&  \frac{a^{2}}{2} \Bigg[  1 + \left(  \frac{\pi \varepsilon}{4} \right)^{2}  \Tilde{\chi}_{1} + \mathcal{H}_{11\ell}
\\
 &+& \mathcal{H}_{12\ell}  \left(  \frac{\pi \varepsilon}{4} \right)^{2}  \Tilde{\chi}_{1} + \mathcal{H}_{13\ell}  \frac{\pi \varepsilon}{4} \sqrt{ \Tilde{\chi}_{1} } \Bigg] \quad .
\end{eqnarray}
where the terms within brackets are the sought corrections due to $\ell \equiv L/\lambda$:
\begin{eqnarray}
\nonumber
\mathcal{H}_{11\ell} &=& \frac{\sin{(4\pi \ell)} }{4\pi \ell} \,\,\, ; \,\,\, \mathcal{H}_{12\ell} =  \frac{\sin{(8\pi \ell)} }{8\pi \ell}   \,\, ,
\\
\mathcal{H}_{13\ell} &=& \frac{3\sin{(2\pi \ell)} +  \sin{(6\pi \ell)}}{3\pi \ell}  \,\, .
\label{eq:H13l}
\end{eqnarray}
These three corrections represent the effects of the shoaling length on the statistics of linear ($\mathcal{H}_{11\ell} $), second-order ($\mathcal{H}_{12\ell} $) and a mixed third-order wave field ($\mathcal{H}_{13\ell} $). The oscillations as a function of $\ell$ are of the same order of magnitude, but differences in the size of first up to third-order terms exist due to a combined effect of steepness and relative water depth. For instance, the second-order $\ell$ oscillation is about $\varepsilon^2 \Tilde{\chi}_1 \sim  20$ larger than the first-order counterpart at the region of highest statistical amplification ($0.3 \lesssim \mu \lesssim 0.6$), while being about $\varepsilon \sqrt{\Tilde{\chi}_1} \sim  5$ larger than the third-order effect.

The averaged potential and kinetic wave energies are defined by eq.~\eqref{eq:energyX0}. For the potential energy, its definition turns out to be half of the approximation used for the variance in eq.~\eqref{eq:variance_app}. Therefore, we have directly:
\begin{eqnarray}
\nonumber
\mathscr{E}_{p} &=&  \frac{a^{2}}{4} \Bigg[  1 + \left(  \frac{\pi \varepsilon}{4} \right)^{2}  \Tilde{\chi}_{1} + \mathcal{H}_{11\ell}
\\
 &+& \mathcal{H}_{12\ell}  \left(  \frac{\pi \varepsilon}{4} \right)^{2}  \Tilde{\chi}_{1} + \mathcal{H}_{13\ell}  \frac{\pi \varepsilon}{4} \sqrt{ \Tilde{\chi}_{1} } \Bigg]
\label{eq:H23l}
\end{eqnarray}
We stress that the variance of FSE being twice the averaged potential wave energy is not a universal result, but a consequence of the choice made above to evaluate the variance in this study. Now we compute the double integral for the kinetic energy $\mathscr{E}_{k}/(a\omega)^2$:
\begin{eqnarray}
\hspace{-0.8cm}
\iint \frac{\textrm{d}x \, \textrm{d}z}{2gL }    \left[ \frac{ \cosh{\varphi} }{ \sinh{ \mu } } \cos{\phi}  +   \frac{3\pi \varepsilon}{4}  \frac{\cosh{(2\varphi)}   }{\sinh^{4}{ \mu }} \cos{(2\phi)} \right]^{2}  
\\
\nonumber
\hspace{-0.8cm}
+ \iint \frac{\textrm{d}x \, \textrm{d}z}{2gL }  \left[ \frac{ \sinh{\varphi} }{ \sinh{ \mu } } \sin{\phi}  +   \frac{3\pi \varepsilon}{4}   \frac{ \sinh{(2\varphi)} }{ \sinh^{4}{ \mu } } \sin{(2\phi)} \right]^{2} .
\end{eqnarray}
Expanding the trigonometric terms leads to:
\begin{eqnarray}
\hspace{-0.9cm}
\nonumber
\mathscr{E}_{k} &=&   \frac{(a\omega)^{2}}{2gL }  \int_{0}^{L}     \textrm{d}x  \Bigg\{  \mathcal{J}_{1} \frac{  \cos^{2}{\phi} }{ \sinh^{2}{ \mu } }   + \mathcal{J}_{12}   \frac{3\pi \varepsilon}{2} \frac{ \cos{\phi} \cos{(2\phi)} }{\sinh^{5}{ \mu }} 
\\
&+& \mathcal{J}_{2}  \left( \frac{3\pi \varepsilon}{4} \right)^{2} \frac{  \cos^{2}{(2\phi)} }{\sinh^{8}{ \mu }}   +   \mathcal{J}_{3} \frac{  \sin^{2}{\phi} }{ \sinh^{2}{ \mu } } 
\\
\nonumber
    &+& \mathcal{J}_{34}   \frac{3\pi \varepsilon}{2} \frac{ \sin{\phi} \sin{(2\phi)} }{\sinh^{5}{ \mu }} + \mathcal{J}_{4}  \left( \frac{3\pi \varepsilon}{4} \right)^{2} \frac{  \sin^{2}{(2\phi)} }{\sinh^{8}{ \mu }}  \Bigg\} , 
\end{eqnarray}
with the set of auxiliary integrals:
\begin{numcases}{}
\mathcal{J}_{1} \equiv \int_{-h}^{0}  \cosh^{2}{\varphi} \, \textrm{d}z = \frac{h}{2} + \frac{ \sinh{(2  \mu )} }{4k}; \,\, \notag\\ \mathcal{J}_{2} \equiv \int_{-h}^{0}  \cosh^{2}{(2\varphi)} \, \textrm{d}z = \frac{h}{2} + \frac{ \sinh{(4  \mu )} }{8k}, \notag\\
\mathcal{J}_{12} \equiv \int_{-h}^{0}  \cosh{\varphi} \, \cosh{(2\varphi)} \, \textrm{d}z = \frac{ 3\sinh{ \mu } +  \sinh{(3 \mu )}}{6k},  \\
\mathcal{J}_{3} \equiv  \int_{-h}^{0}  \sinh^{2}{\varphi} \, \textrm{d}z = - \frac{h}{2} + \frac{ \sinh{(2  \mu )} }{4k}; \, \notag\\ \mathcal{J}_{4} \equiv \int_{-h}^{0}  \sinh^{2}{(2\varphi)} \, \textrm{d}z = - \frac{h}{2} + \frac{ \sinh{(4  \mu )} }{8k}, \notag\\
\mathcal{J}_{34} \int_{-h}^{0}  \sinh{\varphi} \, \sinh{(2\varphi)} \, \textrm{d}z = \frac{  2\sinh^{3}{ \mu }}{3k}.  \notag
\end{numcases}

Ergo, the kinetic energy integrals can be rearranged:
\begin{eqnarray}
\frac{2g \mathscr{E}_{k}}{(a\omega)^{2} } 
&=&\int_{0}^{L}  \frac{\textrm{d}x}{L}  \Bigg\{   \frac{   \mu   \cos{(2\phi)} }{ 2k\sinh^{2}{ \mu } } + \frac{  \sinh{(2  \mu )}  }{ 4k \sinh^{2}{ \mu } }  
\\
\nonumber
&+& \left( \frac{3\pi \varepsilon}{4} \right)^{2}  \frac{   \mu   \cos{(4\phi)} }{ 2k\sinh^{8}{ \mu } } +  \left( \frac{3\pi \varepsilon}{4} \right)^{2}   \frac{  \sinh{(4  \mu )}  }{ 8k \sinh^{8}{ \mu } }
\\
\nonumber
&+&   \pi \varepsilon \cdot \frac{ \sin{\phi} \sin{(2\phi)} }{k\sinh^{2}{ \mu }} 
\\
\nonumber
&+& \left( \frac{\pi \varepsilon}{4} \right) \frac{ \cos{\phi} \cos{(2\phi)} }{k\sinh^{5}{ \mu }} \left(  3\sinh{ \mu } +  \sinh{(3 \mu )}\right)  \Bigg\} \,\, .
\end{eqnarray}
Because the wavenumber and relative water depth vary slowly as compared to $\phi$, we can factor out the term $\big(2k\sinh^{2}{ \mu }\big)^{-1}$ from all integrals. Therefore, considering the dispersion relation $\omega^{2} = gk \tanh{ \mu }$, we have $\big[ (a\omega)^{2}/(2g) \big] (2k\sinh^{2}{ \mu })^{-1} = (a^{2}/4)\big[ \tanh{ \mu } / \sinh^{2}{ \mu } \big] \equiv a^2/(2\sinh{(2 \mu )})$, thus:
\begin{eqnarray}
\nonumber
\hspace{-0.8cm}
 \mathscr{E}_{k} &=&  \frac{a^{2}}{2\sinh{(2 \mu )}} \int_{0}^{L}  \frac{\textrm{d}x}{L}  \Bigg[   \mu   \cos{(2\phi)}  + \frac{  \sinh{(2  \mu )}  }{ 2 }  
 \\
 \nonumber
 \hspace{-0.8cm}
 &+& \left( \frac{3\pi \varepsilon}{4} \right)^{2}  \frac{   \mu  \cos{(4\phi)} }{ \sinh^{6}{ \mu } } +  \left( \frac{3\pi \varepsilon}{4} \right)^{2}   \frac{  \sinh{(4  \mu )}  }{ 4\sinh^{6}{ \mu } } 
\\
\nonumber 
\hspace{-0.8cm}
&+&   2  \pi \varepsilon \cdot  \sin{\phi} \sin{(2\phi)} 
\\
\hspace{-0.8cm}
&+& \left( \frac{\pi \varepsilon}{4} \right) \frac{ 2\cos{\phi} \cos{(2\phi)} }{\sinh^{3}{ \mu }} \left(  3\sinh{ \mu } +  \sinh{(3 \mu )}\right) \Bigg] .   
\end{eqnarray}
\begin{figure*}
\hspace{+0.99cm}
\minipage{0.4\textwidth}
    \includegraphics[scale=0.58]{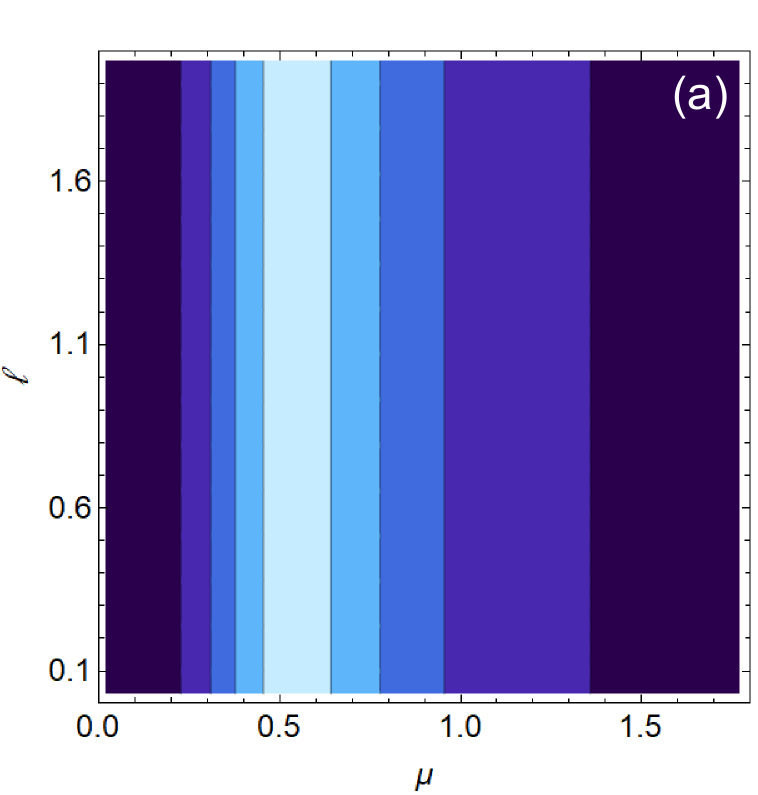}
\endminipage
\hfill
\minipage{0.53\textwidth}
    \includegraphics[scale=0.58]{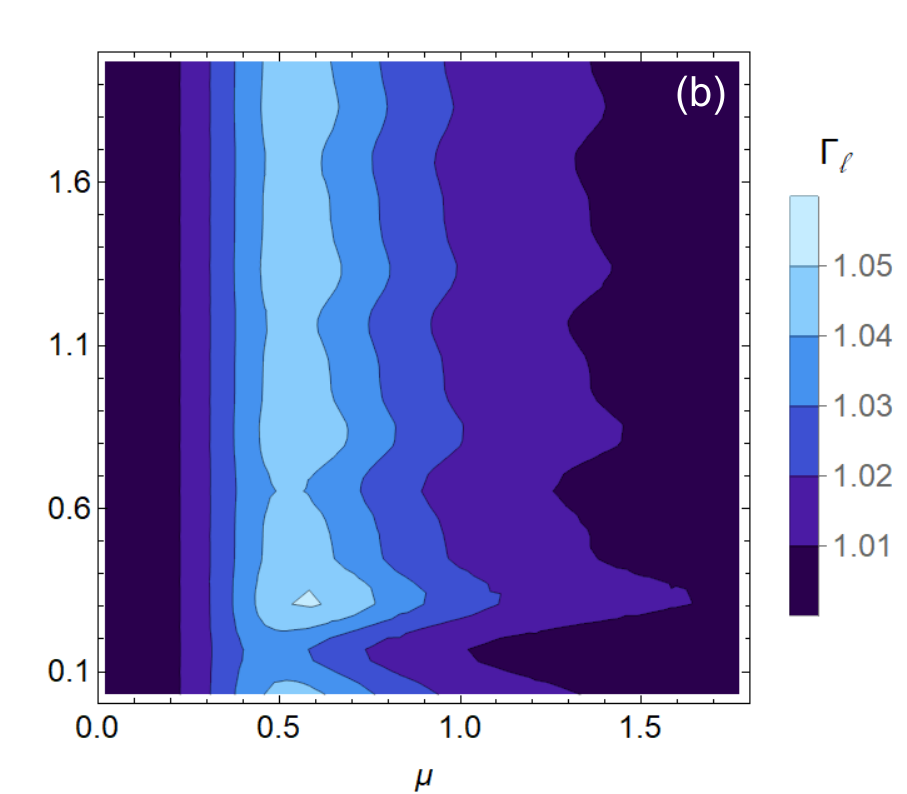}
\endminipage
\caption{Contour plot of the (a) $\ell$-independent $\Gamma$ from eq.~\eqref{eq:Rayexc} and (b) $\ell$-dependent spectral correction $\Gamma_{\ell}$ from eq.~\eqref{eq:Gammal}, as a function of water depth with a fixed wave steepness of $\varepsilon = 1/20$.}
\label{Fig:f3}
\end{figure*}
Further algebra is needed. Once the identity $\big[ 2 / \sinh{(2 \mu )}\big](\sinh{(4 \mu )} / 4 ) = \cosh{(2 \mu )}$ is realized together with the approximation $(9 \cdot 3/2)\sinh^{-6}{ \mu } \approx \chi_{1}$ at the narrow range of the peak in amplification $ \mu  \sim 1/2$, we find:
\begin{eqnarray}
\nonumber
\mathscr{E}_{k} &=&  \frac{a^{2}}{4}  \int_{0}^{L}  \frac{\textrm{d}x}{L}  \Bigg\{  \frac{2  \mu  \cos{(2\phi)}  }{\sinh{(2 \mu )}}   + 1   +  \left( \frac{\pi \varepsilon}{4} \right)^{2}   \chi_{1}   
\\
\nonumber
&+&   \frac{2  \mu }{\sinh{(2 \mu )}}  \left( \frac{\pi \varepsilon}{4} \right)^{2} \left[ \chi_{1} -  \frac{  9  }{2 \sinh^{6}{ \mu } } \right] \cos{(4\phi)}
\\
\nonumber
&+& \frac{4  \pi \varepsilon}{\sinh{(2 \mu )}} \cdot  \sin{\phi} \sin{(2\phi)} 
\\
\nonumber
&+& \pi \varepsilon \cdot \frac{  \left(  3\sinh{ \mu } +  \sinh{(3 \mu )}\right) }{\sinh{(2 \mu )} \sinh^{3}{ \mu }} \cos{\phi} \cos{(2\phi)}    \Bigg\} 
\\
\nonumber
&=& \frac{a^{2}}{4}  \Bigg\{  1   + \left( \frac{\pi \varepsilon}{4} \right)^{2}   \chi_{1} +  \frac{2  \mu  \big( 2\mathcal{I}_{1}-1 \big) }{\sinh{(2 \mu )}}       
\\
\nonumber
&+&    \frac{2  \mu  \big( 2\mathcal{I}_{2}-1 \big) }{\sinh{(2 \mu )}}  \left( \frac{\pi \varepsilon}{4} \right)^{2} \left[ \chi_{1} -  \frac{  9  }{2 \sinh^{6}{ \mu } } \right] +  \frac{4  \pi \varepsilon}{\sinh{(2 \mu )}} \mathcal{I}_{34}  
\\
&+&  \pi \varepsilon  \frac{  \left(  3\sinh{ \mu } \sinh{(3 \mu )}\right) }{\sinh{(2 \mu )} \sinh^{3}{ \mu }} \mathcal{I}_{12}  \Bigg\}.  
\end{eqnarray}
Then, resorting to the set of integrals from eq.~(\ref{eq:setint}) we finally achieve a full expression for the kinetic energy:
\begin{eqnarray}
\mathscr{E}_{k} &=&  \frac{a^{2}}{4} \Bigg\{   1   + \left( \frac{\pi \varepsilon}{4} \right)^{2}   \chi_{1}  
\\
\nonumber
&+& \Big[ \mathcal{H}_{31\ell}  + \mathcal{H}_{32\ell} \left( \frac{\pi \varepsilon}{4} \right)^{2}   \chi_{1}  + \mathcal{H}_{33\ell} \Big]     \Bigg\}, 
\label{eq:A_kine_energy}
\end{eqnarray}
with corresponding corrections due to the shoaling length:
\begin{eqnarray}
\mathcal{H}_{31\ell} &=& \frac{\sin{(4\pi \ell)} }{4\pi \ell} \quad , \quad
\mathcal{H}_{32\ell} = \frac{\sin{(8\pi \ell)} }{8\pi \ell} \quad ,
\\
\nonumber
\mathcal{H}_{33\ell} &=&   \pi \varepsilon\left(  \frac{3\sin{(2\pi \ell)} +  \sin{(6\pi \ell)}}{12\pi \ell} \right) \left(  \frac{    3\sinh{ \mu } +  \sinh{(3 \mu )} }{\sinh{(2 \mu )} \sinh^{3}{ \mu }} \right)
\\
\nonumber
  &+&  \left[ \frac{2  \mu}{\sinh{(2 \mu )}} - 1  \right] \left[ \frac{\sin{(4\pi \ell)} }{4\pi \ell} +  \frac{\sin{(8\pi \ell)} }{8\pi \ell} \left( \frac{\pi \varepsilon}{4} \right)^{2}   \chi_{1}  \right]
\\
\nonumber
&+& \frac{4  \pi \varepsilon}{\sinh{(2 \mu )}} \frac{  \sin^{3}{(2\pi \ell)}}{3\pi \ell} 
\\
\nonumber
&-&  \frac{2  \mu  }{\sinh{(2 \mu )}} \frac{\sin{(8\pi \ell)} }{8\pi \ell} \left( \frac{\pi \varepsilon}{4} \right)^{2} \frac{  9  }{2 \sinh^{6}{ \mu }} \, .
\label{eq:setint2}
\end{eqnarray}
In order to rearrange $\mathcal{H}_{33\ell}$ in the shape as its leading order counterpart $\mathcal{H}_{13\ell}$, we define the ratio $f_{3}$ such that $\mathcal{H}_{33\ell} \equiv    \left(\mathcal{H}_{13\ell}\frac{\pi \varepsilon}{4} \sqrt{ \Tilde{\chi}_{1} }\right)f_3  $:
\begin{eqnarray}
 f_{3}  &=&  \frac{    3\sinh{ \mu } +  \sinh{(3 \mu )} }{\sinh{(2 \mu )} \sinh^{3}{ \mu } \sqrt{ \Tilde{\chi}_{1} }} 
 \\
\nonumber
 &+& \frac{16}{ \sinh{(2 \mu )} \sqrt{ \Tilde{\chi}_{1} } } \frac{  \sin^{3}{(2\pi \ell)}}{3\sin{(2\pi \ell)} +  \sin{(6\pi \ell)}}  
\\
\nonumber
&+& \left[ \frac{2  \mu  }{\sinh{(2 \mu )}} - 1  \right] \frac{3}{\pi \varepsilon \sqrt{ \Tilde{\chi}_{1} }}  \frac{\sin{(4\pi \ell)} }{  3\sin{(2\pi \ell)} +  \sin{(6\pi \ell)} } \times
\\
\nonumber
&{}&  \left[ 1 +  \cos(4\pi\ell) \left( \frac{\pi \varepsilon}{4} \right)^{2}   \tilde\chi_{1}  \right]
\\
\nonumber
&-& \frac{2  \mu  }{\sinh{(2 \mu )}} \frac{\sin{(8\pi \ell)} }{3\sin{(2\pi \ell)} +  \sin{(6\pi \ell)}} \frac{\pi \varepsilon}{4 \sqrt{ \Tilde{\chi}_{1} }}  \frac{  27  }{16 \sinh^{6}{ \mu }}.
\end{eqnarray}
Wherefore, the total energy $\mathscr{E}_{k}+ \mathscr{E}_{p}$ becomes:
\begin{eqnarray}
\nonumber
\mathscr{E} &=&   \frac{a^{2}}{4} \Bigg\{  2 + \left(  \frac{\pi \varepsilon}{4} \right)^{2} \left( \Tilde{\chi}_{1}    +   \chi_{1}  \right) + \Big[ 2\mathcal{H}_{11\ell}  
\\
\nonumber
&+&   \mathcal{H}_{12\ell} \left(  \frac{\pi \varepsilon}{4} \right)^{2} \left( \Tilde{\chi}_{1}    +   \chi_{1}  \right)  +  (1+f_{3}) \mathcal{H}_{13\ell} \frac{\pi \varepsilon \sqrt{ \Tilde{\chi}_{1} }}{4}  \Big] \Bigg\}
\\
\nonumber
&=&   \frac{a^{2}}{2} \Bigg\{ \Big[  1 +  \mathcal{H}_{11\ell} \Big] + \Big[  1 + \mathcal{H}_{12\ell} \Big]  \left(  \frac{\pi \varepsilon}{4} \right)^{2} \frac{ \left( \Tilde{\chi}_{1}    +  \chi_{1} \right)}{2}    
\\
&+&  \frac{(1+f_{3})  }{2} \mathcal{H}_{13\ell} \frac{\pi \varepsilon}{4} \sqrt{ \Tilde{\chi}_{1} } 
\Bigg\}  .
\label{eq:energyFIM}
\end{eqnarray}
As discussed in the paragraph following eq.~(\ref{eq:H13l}), the kinetic counterpart $\mathcal{H}_{13\ell}\frac{\pi \varepsilon}{4} \sqrt{ \Tilde{\chi}_{1} }$ of the third-order shoaling length correction is about 1/5 of its second-order counterpart $\mathcal{H}_{12\ell} \left(  \frac{\pi \varepsilon}{4} \right)^{2} \left( \Tilde{\chi}_{1} + \chi_{1} \right)/2$. Now, because second-order wave fields travelling from deep to intermediate water have bound values of $0.5 < f_{3} < 1.3$, the term $0.75 <(1+f_{3})/2 < 1.15$ is too small to make the third-order effect in $\ell$ comparable to the second-order one. Thus, in both regions (short and long $\ell$ scales) the second-order effect due to $\mathcal{H}_{12\ell}$ has the leading order.
Accordingly, plugging eqs.~(\ref{eq:energyVAR}) and (\ref{eq:energyFIM}) into the non-homogeneous parameter in eq.~(\ref{eq:Rayexc}), we arrive at:
\begin{figure*}
\centering
    \includegraphics[width=0.75\textwidth]{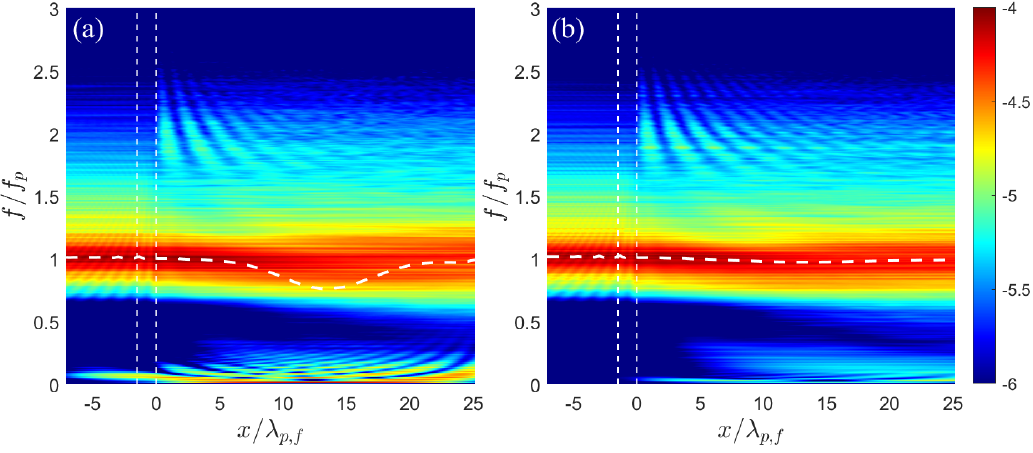}
    \caption{Spatial evolution of the wave spectrum for case 7, simulated with two different extents of the damping zone after $x/\lambda_{p,f}=25$: (a) $L_{\textrm{damp}} = 4$~m $\approx 3 \lambda_{p,f}$ and (b) $L_{\textrm{damp}} = 21$~m $ \approx 20 \lambda_{p,f}$. The colour scale depicts the value of log$_{10}(S(f,x))$ with spectrum $S(f,x)$ in m$^2$/Hz. The vertical white dash lines represent the extent of the plane slope. The horizontal dash line represents the local peak frequency of the wave spectrum.}
\label{Fig:append_spec}
\end{figure*}
\begin{figure*}
\centering
    \includegraphics[width=0.75\textwidth]{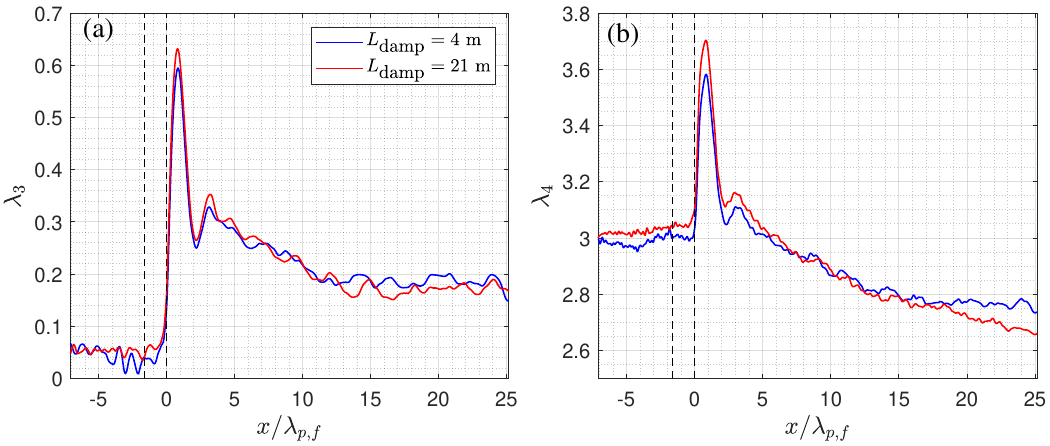}
    \caption{Spatial evolution along the NWF of (a) skewness and (b) kurtosis for case 7, simulated with two choices of relaxation zone lengths.}
\label{Fig:append_stat}
\end{figure*}
\begin{eqnarray}
\hspace{-0.4cm}
\Gamma_{\ell} \approx \frac{ \left[ 1 + \frac{\sin{(4\pi \ell)} }{4\pi \ell} \right]  + \left[ 1 + \frac{\sin{(8\pi \ell)} }{8\pi \ell} \right] \left(  \frac{\pi \varepsilon}{4} \right)^{2} \Tilde{\chi}_{1} }{   \left[ 1 + \frac{\sin{(4\pi \ell)} }{4\pi \ell} \right]  + \left[ 1 + \frac{\sin{(8\pi \ell)} }{8\pi \ell} \right] \left(  \frac{\pi \varepsilon}{4} \right)^{2} \frac{\left( \Tilde{\chi}_{1}    +  \chi_{1} \right)}{2}  } \quad .
\label{eq:Gammal}
\end{eqnarray}
Oscillations up to second order in $\ell$ can be seen in figure~\ref{Fig:f3}(b) over a wide range of relative water depth. In shallow water, the oscillations are negligible, whereas in the region of rogue wave amplification $\mu \sim 1/2$ are small. Oscillations start to grow for $\mu \ge 1/2$ all the way to deep water, but $\Gamma - 1$ vanishes for large $\mu$ so the impact of the oscillations is limited. 

\section{Effects of low-frequency waves  \label{sec:append_LF}}
Low-frequency (LF) sub-harmonics often occur as a result of nonlinear wave-wave interactions during nonlinear simulations. Particular attention should be paid to LF wave damping while running simulations with long duration, as LF wave energy may accumulate and affect the primary harmonics in return. Conventionally, in W3D, a damping zone of three characteristic wavelengths at the end of the computational domain works well in absorbing waves. However, in the present work, as the duration is long and the nonlinear wave-wave interaction is quite active after the shoal, taking $L_{\textrm{damp}} = 4$~m $\approx 3 \lambda_{p,f}$ is not sufficient. Rather, the length of the damping zone should be set according to the LF wavelength, i.e., $L_{\textrm{damp}} = 21$~m $ \approx 20 \lambda_{p,f} \approx 3 \lambda_\textrm{LF}$, with the LF waves having a characteristic frequency $f_\textrm{LF} \approx 0.05$~Hz. 

Figure~\ref{Fig:append_spec} displays the spatial evolution of the wave spectrum for case 7 with $L_{\textrm{damp}} \approx  3 \lambda_{p,f}$ (in panel a) and $L_{\textrm{damp}} \approx  20 \lambda_{p,f}$ (in panel b). The vertical dash lines indicate the range of the shoal, and the horizontal dash lines represent spectral peak frequency evaluated as $f_p \equiv \left( \int_{0}^{f_{\textrm{Nyq}}} fS^4(f) \textrm{d}f \right) / \left( \int_{0}^{f_{\textrm{Nyq}}} S^4(f) \textrm{d}f \right) $ with $f_{\textrm{Nyq}}=50$~Hz according to Nyquist sampling theorem. Clearly, with a short damping zone, the LF components receive a considerable amount of energy during the simulation and affect markedly the peak frequency atop and after the shoal in panel (a). With the longer damping zone, however, as shown in panel (b), the LF components are effectively suppressed. Consequently, the longer damping zone $L_{\textrm{damp}} \approx  20 \lambda_{p,f}$ was adopted in all simulations presented in this study.

Moreover, with these two choices of $L_{\textrm{damp}}$, the effect of LF components on statistical parameters $\lambda_3$ and $\lambda_4$ can be discussed. In figure~\ref{Fig:append_stat}, the spatial evolutions of $\lambda_3$ and $\lambda_4$ are displayed. It is observed that strong LF components (i.e., with the reduced $L_{\textrm{damp}}$) result in lower levels of both parameters after the shoal. This means that, in nature, the LF components of coastal waves (possibly reflected from the shoreline or released during depth-induced wave breaking) could play a role in the mitigation of rogue wave risk.

\bibliography{Maintext}

\end{document}